\documentclass[twocolumn,amssymb, nobibnotes, floatfix, aps, prl]{revtex4-1}
\usepackage{hyperref}
\usepackage[utf8]{inputenc}
\usepackage{amssymb}
\usepackage{amsmath}
\usepackage{multirow}
\usepackage{graphicx}
\graphicspath{{./fig/}}
\usepackage{makecell}
\usepackage{array}
\usepackage{booktabs}
\usepackage[export]{adjustbox}
\usepackage{color}
\usepackage{physics}
\usepackage{marginnote}
\usepackage{pdfcomment}

\renewcommand{\vec}[1]{\boldsymbol{#1}}

\begin{document}


\title{Edge modes and non-local conductance in graphene superlattices}


\author{{Rory Brown}$^1$}
\email{rory.brown-3@postgrad.manchester.ac.uk} 
\author{Niels R. Walet$^1$}
\email{Niels.Walet@manchester.ac.uk}
\author{Francisco Guinea$^{1,2}$}
\email{Francisco.Guinea@manchester.ac.uk}
\affiliation{$^1$School of Physics and Astronomy, University of Manchester, Manchester, M13 9PY, UK}
\affiliation{$^2$Imdea Nanoscience, Faraday 9, 28015 Madrid, Spain}

\date{\today}

\begin{abstract}
We study the existence of edge modes in gapped Moir\'e superlattices in graphene monolayer ribbons. We find that the superlattice bands acquire finite Chern numbers, which lead to a Valley Hall Effect. The presence of dispersive edge modes is confirmed by calculations of the band structure of realistic nanoribbons using tight binding methods. These edge states are only weakly sensitive to disorder, as short-range scattering processes lead to mean free paths of the order of microns. The results explain the existence of edge currents when the chemical potential lies within the bulk superlattice gap, and offer an explanation for existing non-local resistivity measurements in graphene ribbons on boron nitride.
\end{abstract}

\pacs{???}

\maketitle


{\it Introduction.} 
Even before the characterization of monolayer graphene, it was proposed that semi-infinite ribbons constructed from a honeycomb lattice of carbon atoms supported electronic states localized at the edges \cite{FWNK96,NFDD96}. Many boundary conditions lead to a flat dispersionless band of edge modes \cite{AB08}, although some boundary conditions such as the armchair edge do not \cite{BF06,BM87}. When additional terms in the Hamiltonian, such as a hopping between second nearest neighbor sites \cite{PGN06} or electron interactions \cite{WSSG08} are considered, these bands become weakly dispersive. The existence of an approximately flat band suggests the presence of magnetism at the edges of graphene, a topic already considered before the characterization of graphene \cite{WFHS99}, see also \cite{NGPNG09,KUPGN12}. Nevertheless, experimental evidence of the existence of these states has remained elusive \cite{Tetal11}, although recent results suggest the presence of electronic transport along the edges of graphene ribbons \cite{Aetal16, Wang16}.

As discussed below, we are interested in edge transport in insulating single layer graphene superlattices which show a gap at the Dirac point, as is observed experimentally for graphene on hexagonal boron nitride (BN). As mentioned above, edge states are present in simple models of gapped graphene, although they do not lead to edge transport as these bands are flat and separated by a gap. The situation is slightly more complicated in gapped bilayer graphene, where the edge bands are dispersive \cite{Cetal08}: here the edges are metallic even when the bulk is insulating. The effects of the Moir\'e superlattice in twisted bilayer graphene have been studied previously \cite{Weckbecker16,Bistritzer11}, and will not be covered in depth here. Recent numerical calculations suggest that dispersive edge modes are responsible for non-local resistivity measurements in single layer graphene on a BN substrate \cite{Metal17}; we shall demonstrate that this is a generic feature of Moir\'e superlattices in single layer graphene, which can be attributed to a Valley Hall Effect.

The existence of flat bands in semimetallic and gapped single layer graphene can be justified on topological grounds \cite{HV11,HKV11}. Band crossings and gapless edges are typical of topological insulators \cite{HK10,LZ11}. A topological insulator is characterized by its Chern numbers, invariant integrals over the bands in the Brillouin Zone. Surface modes within a given gap arise when these numbers are finite.
It can be shown that the non-trivial gapless edge modes in bilayer graphene can be understood if each valley is considered an independent electronic system: the integrated Chern number for each valley allows us to define these valleys as topological insulators \cite{LMBM11}. We note that the integrated Chern number  per valley in simple models of gapped single layer graphene is a half-integer, instead of an integer as in a topological insulator \cite{Watanabe10}. The existence of a finite Berry curvature is not sufficient to turn each valley into a topological insulator, and thus explains the persistence of the gap in a monolayer, despite the existence of edge modes. For the case of gapped bilayer graphene, the existence of gapless edge modes is experimentally supported by a number of non-local transport measurements \cite{Setal15,Setal15b,Jetal15,Letal16,Zetal17}.

\begin{figure}
	\includegraphics[height=0.27\textheight]{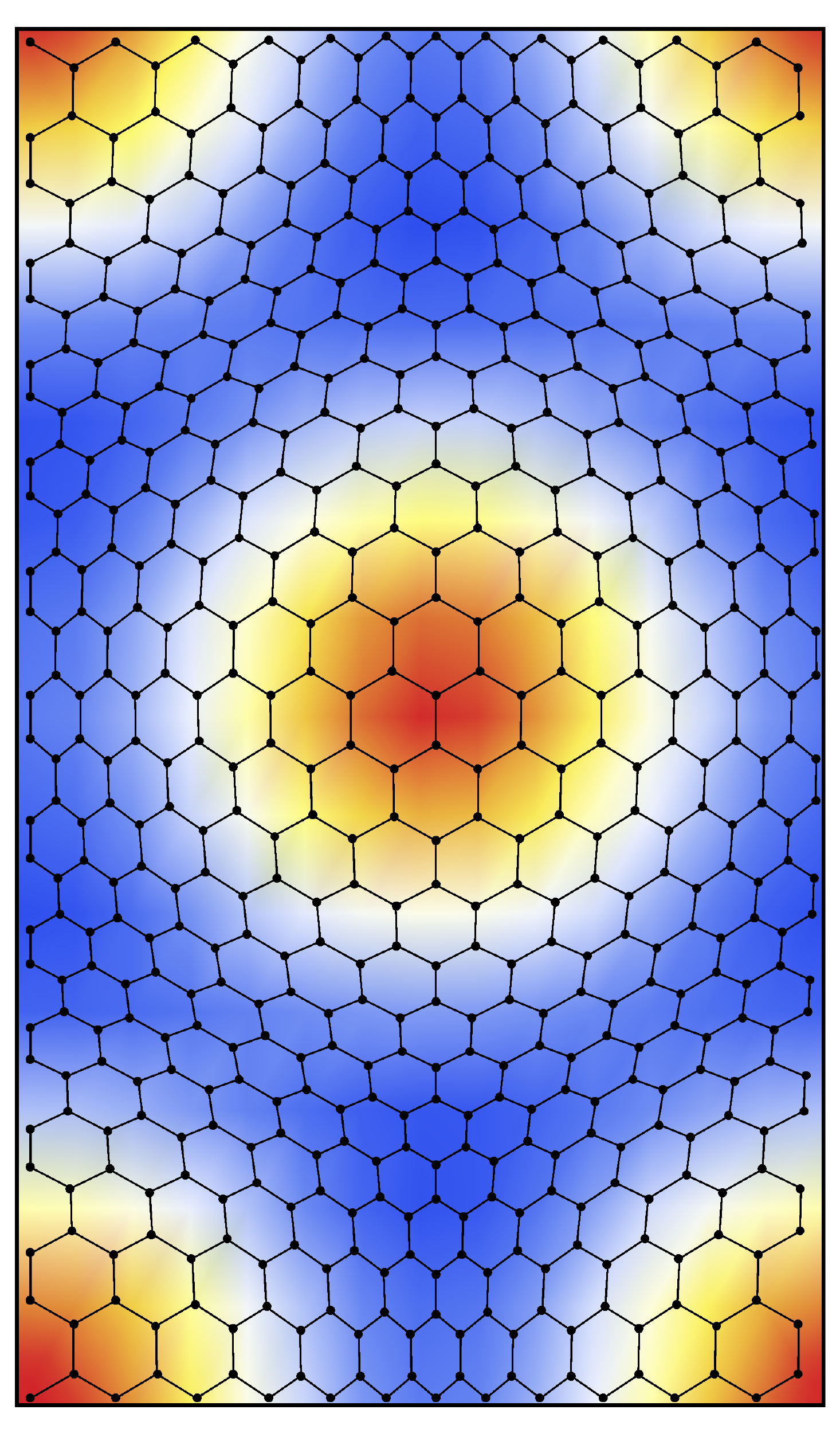}
	\includegraphics[height=0.27\textheight]{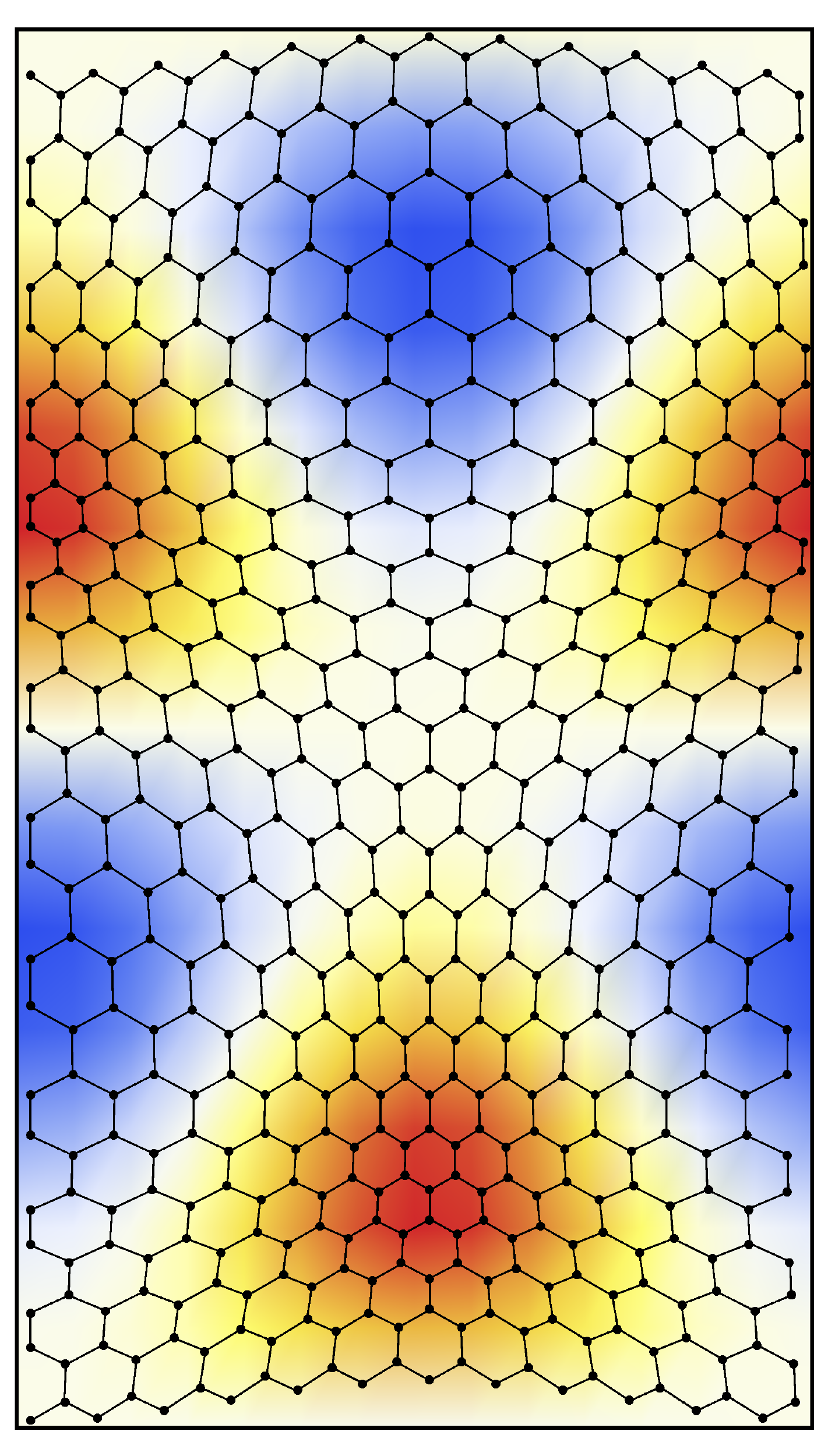}
	\caption{Schematic representation of the even (left) and odd (right) parity scalar and vector potentials; each potential has an equivalent strength of 0.05 eV. The color scale gives the value of the local scalar potential. The vector potential describes changes in the local bond lengths due to the lattice mismatch with the substrate;
lattice deformations are exaggerated by a factor of 10 for illustration purposes. Other sources of a gauge potential are also possible \cite{Jetal17}.
For details of the gauge potential, see \cite{SUP}.
\label{fig:superlattice}}
\end{figure}

A mechanism which leads to gapped monolayer graphene involves breaking its inversion symmetry through the formation of a commensurate superlattice using graphene on BN, where the slight lattice mismatch leads to large-unit-cell superlattices \cite{Yetal11,Petal13,Detal13,Hetal13,Wetal14,Yetal14}. Typically, these superlattices have a unit cell built from about $50 \times 50$ graphene unit cells. Gapped single layer graphene superlattices show strong evidence of electronic edge transport \cite{Getal14,Zetal17} (see also \cite{Cetal12,Aetal16}). It has been suggested that in addition to edge modes, topological currents derived from bulk states can contribute to the observed valley currents \cite{LSSL15}. In this Letter we shall demonstrate that interactions with the BN substrate produce edge modes that are dispersive in the vicinity of the Dirac points and have energies within the bulk gap, turning graphene superlattices into valley Hall insulators. The resulting edge modes act as conduction paths where the gap would otherwise prevent electronic transport. The bulk electronic structure of graphene superlattices is reasonably well understood. If the resulting superlattice has a long wavelength compared to the bond lengths, the bonds in graphene undergo only a slight distortion. We can describe this in terms of a periodic interaction  depending on only six parameters (seven if a constant gap is included). This means that we only consider coupling via the first harmonic functions of the superlattice, which is sufficient given the large size of the superlattice unit cell compared to that of graphene \cite{SUP}. A number of models can be used to obtain reasonable values for these parameters \cite{KUM12,SSL13,SGSG14,JRQM14,NP14,MK14,Jetal17}. 

In the following, we analyze the formation of edge states in monolayer graphene superlattices. We show, using topological arguments, that the superlattice potential induces non-trivial integer Chern numbers which turn each valley in monolayer graphene into an effective topological insulator. This is supported by extensive numerical calculations of the electronic structure of graphene nanoribbons, which show that edge states are a generic feature of these superlattices. As a result, graphene nanoribbons with a superlattice potential become effectively gapless even if the bulk shows a significant gap, $\Delta_\text{bulk} \approx 50 - 100\,\mathrm{meV}$. We define the local current operator and find that the resultant current distribution along the width of the graphene nanoribbons is in agreement with experimental results \cite{Zetal17}. Finally, we estimate that the localization length due to edge disorder is sufficiently large for edge transport to contribute to measurements in graphene devices.

{\it The model. Superlattice bands and Berry curvature.}
The band structure and electronic Berry curvature of graphene superlattices can be analyzed using the Dirac equation, where the superlattice potential is included by diagonalisation in a  truncated basis of unperturbed plane waves, while ensuring that the results converge with respect to the basis size truncation, as explained below. The superlattice potential includes the seven terms mentioned above. Apart from the uniform constant gap, these are	two scalar potentials, mass gaps, and gauge fields, each with even or odd parity (denoted as $V_s^{e,o}, V_{\Delta}^{e,o}, V_g^{e,o}$), see \cite{WPMGF13}. We neglect the small intervalley coupling terms. The modulation of the six potentials only depends on the first star of reciprocal lattice vectors, $\vec{\bf G}$. Here a star is made up of equivalent vectors in the reciprocal lattice, and each subsequent star is further removed from the origin \cite{SUP}. The first star includes the six reciprocal lattice vectors which generate the even and odd first harmonic functions of the superlattice. A schematic representation of the effect of these perturbations is shown in Fig.~\ref{fig:superlattice}.

The electronic structure of the superlattice is determined by diagonalization of the Hamiltonian in a basis of momentum wavefunctions,
\[| \psi_{\vec {\bf k}}\rangle  = \sum_{i=1, \cdots , n_{\text{max}}} \sum_{j=1, \cdots , n_i} \alpha_{i,j} \left| \vec{\bf k} +  m_{ij}  \vec{\bf G}_j \right\rangle_{e,h}, \]
where $\vec{\bf k}$ is a wavevector within the superlattice Brillouin Zone, $e , h$ denote electron and hole states, the set $\{ \vec{\bf G}_j \}$ with $j = 1 , \cdots , 6$ is the first star of reciprocal lattice wavevectors, and the vectors $ m_{i,j}  \vec{\bf G}_j$ label the $j^{th}$ vector in the $i^{th}$ star of reciprocal lattice vectors, which includes $n_i$ vectors equivalent by symmetry. The degeneracies of the first 5 stars are $n_{i=1,\cdots , 5} = \{ 6 , 6 , 6 , 12 , 6 \}$. The total number of states included in the Hamiltonian for $n_{\text{max}} = 1 , \cdots , 5$ is $n_{\text{states}} = \{ 14 , 26, 38 , 62 , 74 \}$. The number of superlattice subbands that we obtain is equal to $n_{\text{states}}$. The diagonalization of the Hamiltonian allows us to determine the coefficients $\alpha_{i,j}$ and the eigenstate $| \psi_{\vec {\bf k}} \rangle$. We have checked that the results for the subbands near the Dirac energy have converged as function of $n_{\text{max}}$.

Once the eigenenergies and eigenvectors of a given band are calculated we numerically evaluate the Berry curvature at discrete points $\vec{\bf k}$ of the superlattice Brillouin Zone, as well as the Chern number, the integrated Berry curvature, using the method by Fukui \emph{et al} \cite{FHF05}. The Berry curvature for the subbands of a graphene superlattice with the superlattice perturbations ($V_s^{e,o}, V_{\Delta}^{e,o}, V_g^{e,o}$) and a constant average gap ($\delta$) is shown in Fig. \ref{fig:berry}. These results are for the experimentally relevant case of a commensurate superlattice with aligned graphene/BN crystallographic axes and a periodicity of 50 graphene unit cells; we have verified that the existence of finite Chern numbers in the superlattice subbands is not sensitive to the choice of physically reasonable superlattice parameters. The Chern number typically changes from one subband to the next. If each valley is considered separately this behavior describes a topological insulator, the number of edge states within a given gap equal to the sum of the Chern numbers of the occupied bands. This system is usually called a valley Hall insulator. A particular valley gives rise to one or more one dimensional bands which circulate around the edges of the sample with a single orientation. The other valley induces a band circulating with opposite chirality.

\begin{figure}
\begin{tabular}{ c c }
 \multirow{3}{*}{{\sffamily(a)~}\includegraphics[height=0.3\textheight,valign=t,trim={0cm 2cm 0 2cm},clip]{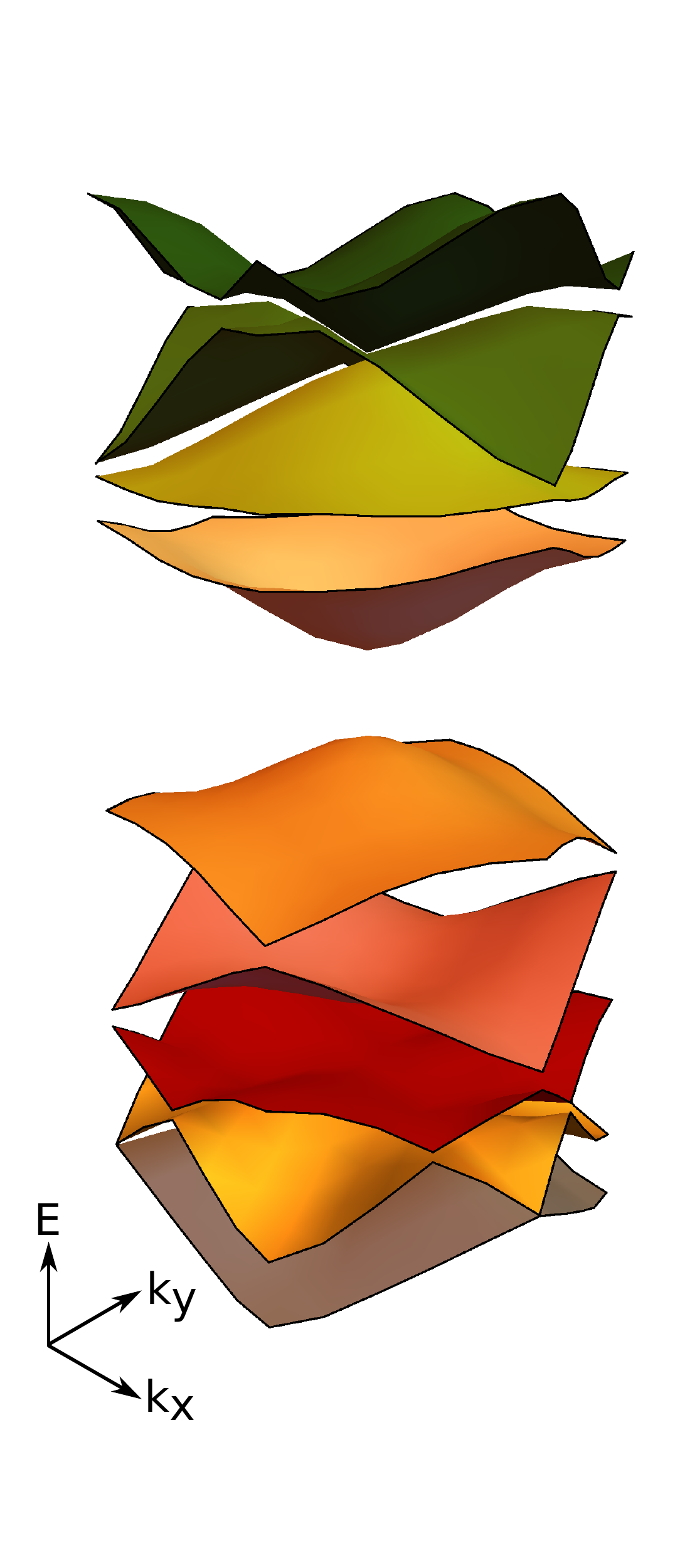}~} &{\sffamily(b)}\includegraphics[height=0.14\textheight,valign=t]{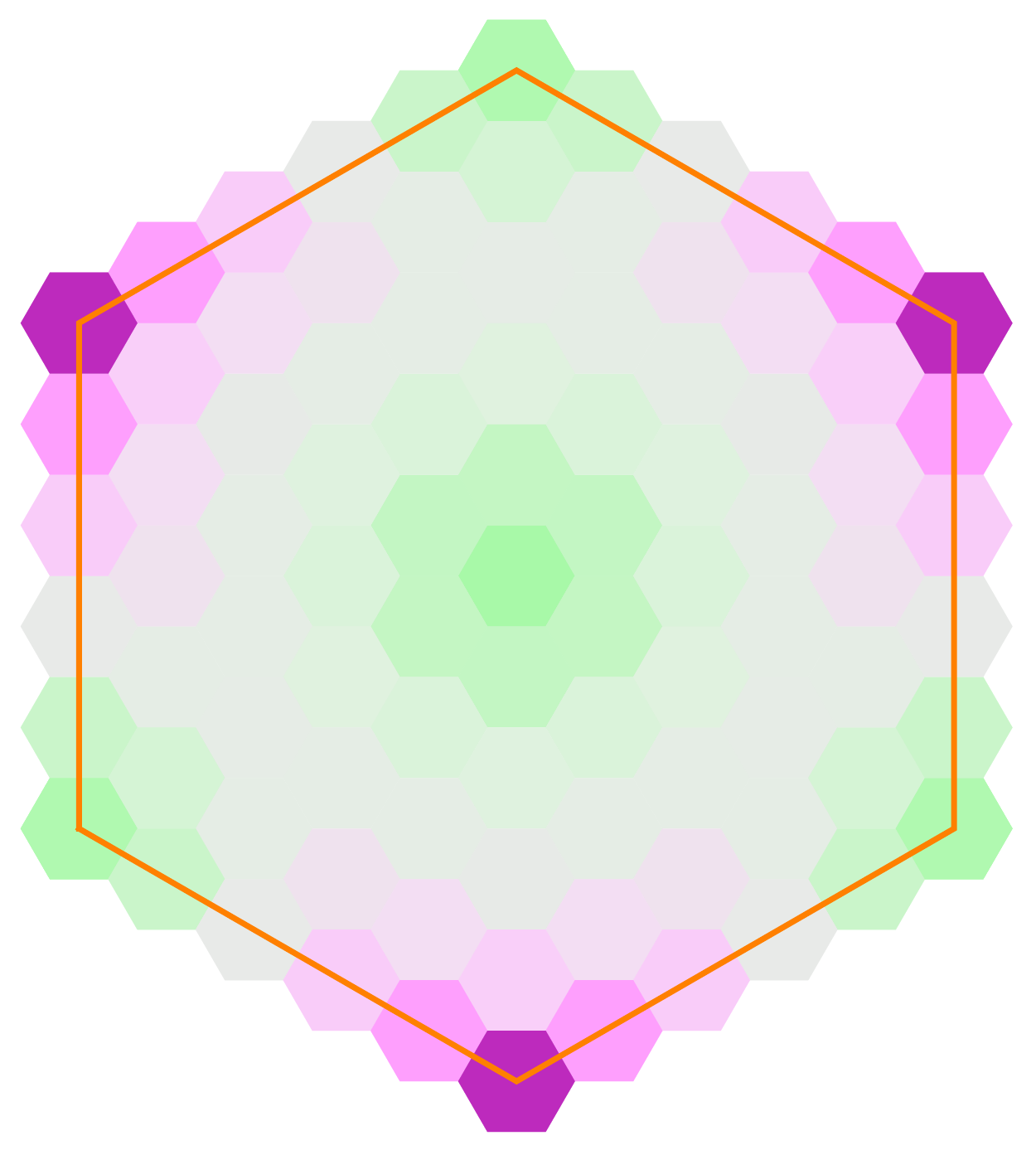} \\
 & {\sffamily(c)}\includegraphics[height=0.14\textheight,valign=t]{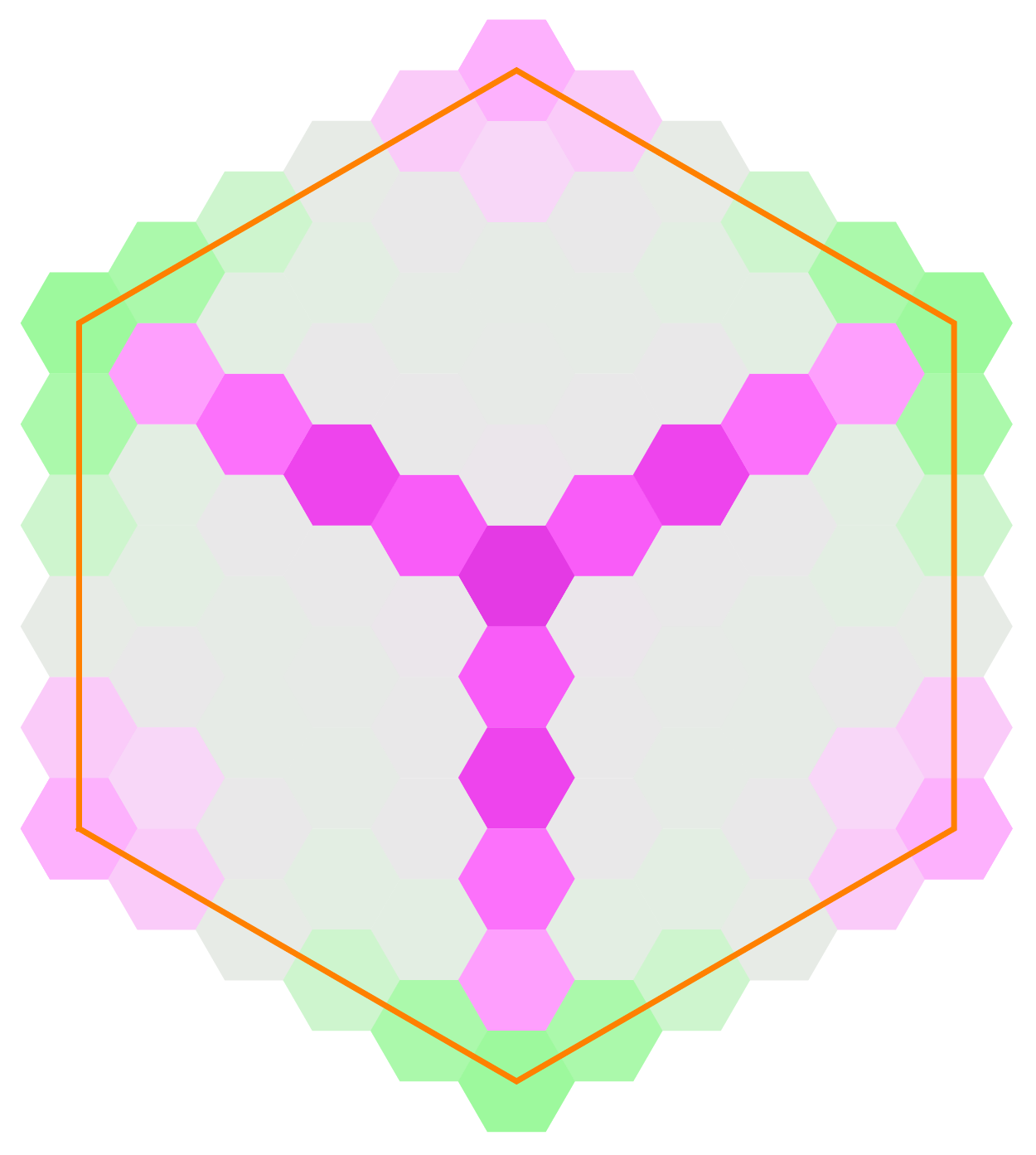}\\&\hspace{0.1in}
\includegraphics[width=0.145\textwidth]{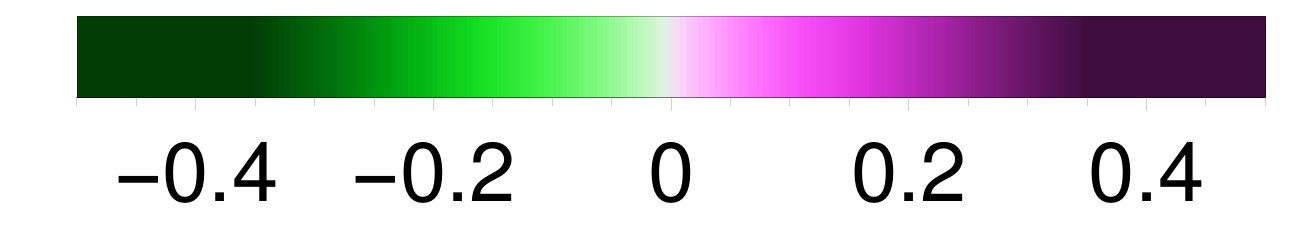}
\end{tabular}
\caption{(a) Subbands of a 50 $\times$ 50 unit cell superlattice, with all perturbation potentials and an average unmodulated gap $\delta$, $(V_s^e, V_s^o, V_{\Delta}^e, V_{\Delta}^o, V_g^e, V_g^o, \delta) = (21, 38, 6, 0, -42, -21, 50)\,\mathrm{meV}$. The Chern densities for the conduction band (b) and first excited band (c) add up to Chern numbers 0 and 1, respectively.}
\label{fig:berry}
\end{figure}

{\it Superlattice nanoribbons.}
Having established that the non-trivial Berry phases in the subbands lead to topologically insulating behaviour, we now analyze the electronic properties of graphene nanoribbons with a similar superlattice structure. We use a real space  tight-binding model, and assume a triangular superlattice. Unit cells containing up to  $48 \times 48$ graphene unit cells, a realistic size, have been considered; smaller superlattice unit cells give qualitatively similar results \cite{SUP}. In the case of the largest cells, the nanoribbon width accommodates only 3 supercells. We have checked that this is sufficient to obtain well converged results, the nature of the edge states no longer changing when adding further	cells. The superlattices considered include an uniform gap and six parameters to describe the superlattice modulation \cite{WPMGF13}. We take the values of these parameters to first order to be the same as in our continuum model, 
and consider zigzag nanoribbons as this is the generic boundary condition describing straight edges \cite{AB08}. For a superlattice with $3 n \times 3 n$ graphene unit cells in the supercell the number of bands at the edge is $4 \times n$, occupying the entire Brillouin zone parallel to the nanoribbon direction, $0 \le k_\parallel d \le 2 \pi$, where $d$ is the length of the nanoribbon unit cell. The bulk $K$ and $K'$ points correspond to $k_\parallel d = 0$.

\begin{figure*}
	\includegraphics[width=0.48\textwidth]{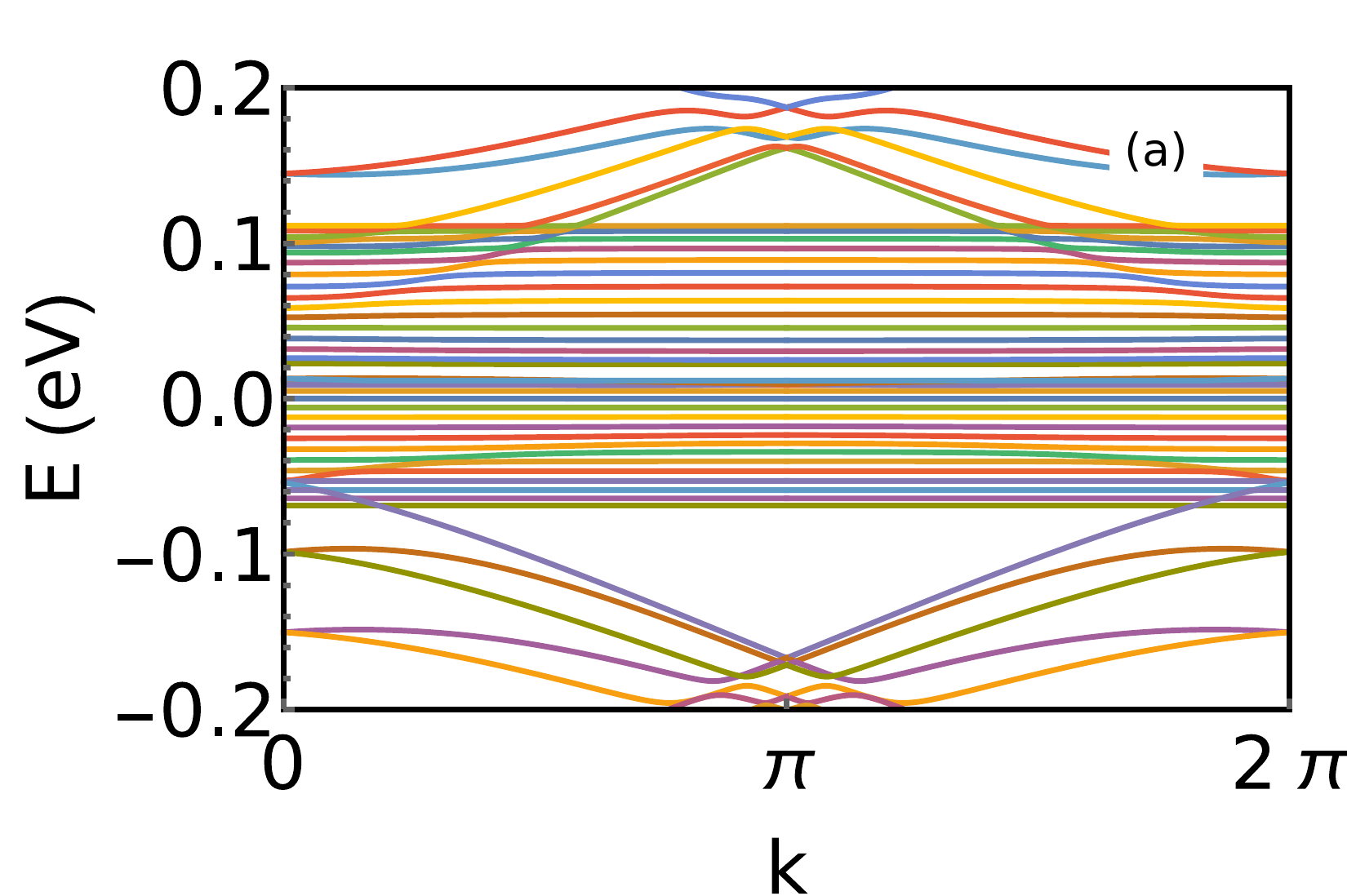}
	\includegraphics[width=0.48\textwidth]{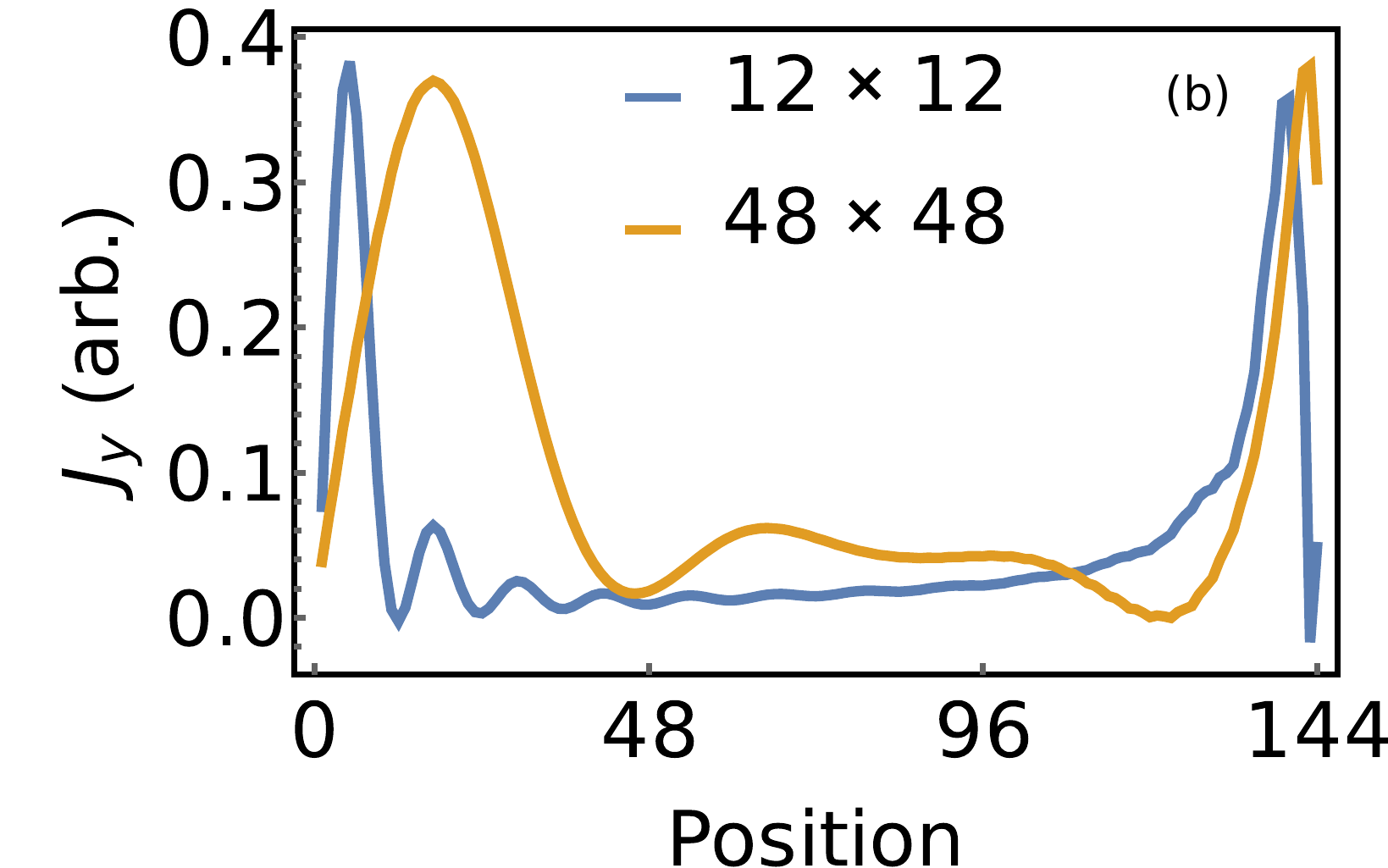}
	\caption{(a) The band structure for nanoribbons 144 lattice sites in width, with a superlattice unit cell of 48 $\times$ 48 graphene unit cells. Flat bands localized at the nanoribbon edges lie within the average bulk gap, and produce a peak in the density of states. Most importantly, they provideconduction channels for low-energy edge transport. (b) The current distribution across nanoribbons of superlattice cell size 12 $\times$ 12 and 48 $\times$ 48 graphene unit cells. Assuming a small voltage and a midgap bias, only bands with energy $E_n$ in the ranges -0.1 $\leq E_n \leq$ 0.1 eV and -0.05 $\leq E_n \leq$ 0.05 eV, respectively, are included. In both figures the superlattice parameters are the same as in Fig. \ref{fig:berry}.}
\label{fig:nanoribbon}
\end{figure*}

Typical results are shown in Fig.~\ref{fig:nanoribbon}. For most combinations of parameters we find edge modes that are dispersive near the Dirac points, with multiple crossings; these tend to be near $k_\parallel d = 0$, and the bands involved are localized at opposite edges. The bands fill a significant fraction of the bulk gap: 
for instance, for the example given in Fig.~\ref{fig:nanoribbon}, of all the states with energies ranging from $-0.5 \leq E \leq 0.5\,\text{eV}$, 28\%  lie in the gap region ($-0.1 \leq E \leq 0.1\,\text{eV}$); see also \cite{SUP}.

While the folding of the bands of graphene with a uniform gap leads to dispersionless bands at the edges of the gap, the velocity associated to the dispersive bands in the presence of a superlattice can be close to the bulk Fermi velocity $v_\text{F}$. These results are consistent with the formation of a topological valley insulator. Note that our choice of superlattice cell size makes possible intervalley scattering at the edges, although the results suggest that it must be weak.

We have modelled the current distribution for cases when the Fermi energy, which varies along the ribbon due to an applied potential, intersects a dispersive band. We describe the current going through a bond between an orbital at position $\vec{\bf r}_i$ and a neighboring orbital at position $\vec{\bf r}_j$ as $t_{ij} {\rm Im} \langle n , k_\text{F}  | c^\dag_i c_j | n , k_\text{F} \rangle$, where $| n , k_\text{F} \rangle$ describes the wavefunction of the band $n$ at the momentum $k_\text{F}$ where it intersects the Fermi energy. Typical results are shown in Fig. \ref{fig:nanoribbon}b. They agree qualitatively with the experiments reported in \cite{Zetal17}; this is generally true irrespective of the details of the superlattice parameters \cite{SUP}.

{\it Effects of disorder.}
We expect the edge states to be localized by disorder. The previous analysis shows that the surface bands are similar to those of a valley Hall insulator near the projection of the $K$ and $K'$ points; near these points the bands disperse with a velocity of the order of the Fermi velocity, $v_{\text{1D}} \approx v_\text{F}$. Backward scattering is only possible through short-range defects. If the range of these defects is of the order of the lattice spacing, $a$, and their concentration is $c_{\text{2D}}$, the contribution of this short-range scattering to the two dimensional inverse scattering time is
\begin{align}
\frac{\hbar}{\tau_{\text{2D}}^{\text{sr}}} &\approx \bar{V}_{\text{2D}}^2 a^2 c_{\text{2D}} D_{\text{2D}} ( \epsilon_\text{F}^{\text{2D}} ) \approx \frac{\bar{V}_{\text{2D}}^2 a ^4 c_{\text{2D}} k_\text{F}^{\text{2D}}}{\hbar v_\text{F}}
\end{align}
where $\bar{V}$ is the strength of the potential created by a single defect and $D_{\text{2D}} ( \epsilon )$ is the density of states.

The decay length of the edge states  $\xi$ must be much larger than $a$, so that the matrix element which describes scattering at the edge by a single impurity is reduced by a factor $( a / \xi )^2$ \cite{Aetal07}.
The typical size of the decay length is $\xi \gtrsim ( \hbar v_\text{F} ) / \Delta$, where $\Delta$ is the gap within which the edge modes are defined. For $\hbar v_\text{F} \approx 6\, \mathrm{eV\,  \times \AA }$ and gaps of order $\Delta \lesssim 0.1\,\mathrm{eV,}$ we find $\xi \gtrsim 10^2\,\mathrm{\AA}$. The inverse scattering time for the edge states is
\begin{align}
\frac{\hbar}{\tau_{\text{1D}}} &\approx \frac{a^2}{\xi^2}  \frac{\bar{V}^2 a^2 c_{\text{2D}} \xi}{\hbar v_{\text{1D}}} + \frac{a^2}{\xi^2} \frac{\bar{V}_{\text{edge}}^2 a^2 c_{\text{edge}}}{\hbar v_{\text{1D}}}  \approx \nonumber \\ &\approx \frac{\bar{V}^2 a^4 c_{\text{2D}}}{\xi \hbar v_\text{F}} + \frac{\bar{V}_{\text{edge}}^2 a^3}{\xi^2 \hbar v_\text{F}}  ,
\end{align}
where we have introduced short range defects characterized by an energy $\bar{V}_{\text{edge}}$ and a concentration $c_{\text{edge}} \approx a^{-1}$ at the edge. The order of magnitude of $k_\text{F}^{\text{2D}}$ at the scales where the edge modes are defined is $k_\text{F}^{\text{2D}} \approx \xi^{-1} \approx \Delta / ( \hbar v_\text{F} )$, so that we obtain 
\begin{align}
\frac{\hbar}{\tau_{\text{1D}}} &\approx \frac{\hbar}{\tau_{\text{2D}}^{\text{sr}}} + \frac{\bar{V}^2_{\text{edge}} \Delta^2 a^3}{( \hbar v_\text{F} )^3}.
\end{align}
The contribution of bulk short range scattering to the transport time is small \cite{Cetal14}, as the mean free path is mostly determined by long range scattering processes. For $a \approx 1$ \AA , 
$\Delta \lesssim 100$ meV, and $\bar{V}_{\text{edge}} \approx 1$ eV, we obtain a mean free path for the edge channels
\begin{align}
\ell_{\text{1D}} = v_{\text{1D}} \tau_{\text{1D}} \approx \frac{( \hbar v_\text{F} )^4}{\bar{V}_{\text{edge}}^2 \Delta^2 a^3} \gtrsim 1 \mu{\rm m}.
\end{align}
The localization length at the edge is approximately given by $\ell^{\text{loc}}_{\text{edge}} \approx \ell_{\text{1D}} N_{\text{ch}}$, where $N_{\text{ch}}$ is the number of channels. In the present case, each edge mode which crosses the Fermi level defines a channel. Even for $N_{\text{ch}} \sim 1$, we obtain a localization length compatible with the dimensions of the samples described in \cite{Zetal17}.

{\it Conclusions.}
We have analyzed the nature of edge states in gapped graphene superlattices. We have shown that non-trivial patterns of Berry curvature are induced in the superlattice Brillouin Zone, giving rise to Chern numbers which are typically non-zero and change from subband to subband. The precise value of these numbers depends on details of the superlattice potential, although they are generally present provided that physically reasonable superlattice parameters are used. The existence of finite Chern numbers in the superlattice bands leads to a Valley Hall Effect. These results are confirmed by real space calculations for superlattice ribbons. We find dispersive bands and crossings near the corners of the Brillouin Zone.

Currents along the superlattice edges are degraded by short-range intervalley scattering, whereas in clean graphene samples electronic transport is limited by long range, intravalley scattering. The effect of disorder localized at the edges is suppressed by the long decay length of the states, due to the small size of the gaps. Simple estimates of the mean free path and localization length associated to edge disorder gives values in the order of microns. This provides an explanation for the low resistivities found in electronic transport measurements of graphene on BN \cite{Getal14}, (see also \cite{Metal17}). Furthermore we have demonstrated that the superlattice is of importance to the transport properties of graphene on a substrate such as BN, or as a means to measure the Valley Hall Effect.

\begin{acknowledgments}
We would like to thank M. Ben Shalom, V. Fal'ko, A. Geim and J. Walbank for useful conversations.
This work was supported by funding from the European Union through the ERC Advanced Grant NOVGRAPHENE through grant agreement Nr. 290846, and from the European Commission under the Graphene Flagship, contract CNECTICT-604391. 
\end{acknowledgments}
\bibliography{edge_states}

\pagebreak
\appendix

\onecolumngrid

\setcounter{equation}{0}
\setcounter{figure}{0}

\renewcommand{\theequation}{A\arabic{equation}}
\renewcommand{\thefigure}{A\arabic{figure}}

\renewcommand{\bibnumfmt}[1]{[#1]}
\renewcommand{\citenumfont}[1]{#1}

\newpage
\onecolumngrid

\title{Supplementary material for ``edge modes and non-local conductance in graphene superlattices"}
\maketitle
\section{Superlattice Hamiltonian}\label{sec:continuum}

\begin{figure}[b]
	\includegraphics[width=0.3\textwidth]{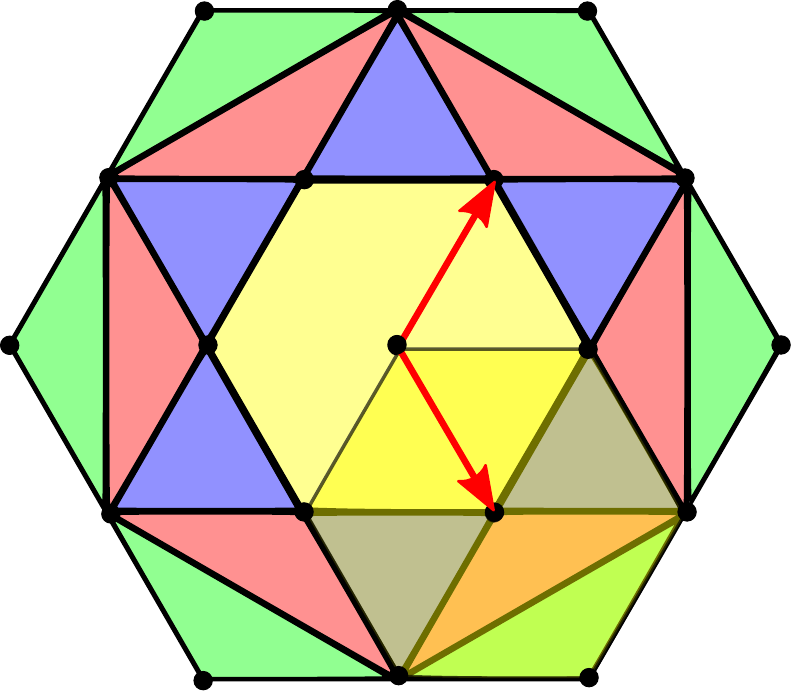}
	\caption{The Brillouin zones and first star of reciprocal lattice vectors for the superlattice. The triangular superlattice formed by two commensurate honeycomb lattices has a hexagonal first Brillouin zone, shown here (\textit{yellow}) with the superlattice reciprocal lattice vectors (\textit{red}). The forms of the second, third and fourth Brillouin zones are shown in blue, red and green respectively. The `first star' refers to the six smallest reciprocal lattice vectors of the superlattice that couple a site to its six nearest neighbors; these neighbors themselves couple to their nearest neighbors via the first star, for $n_{\text{max}}$ iterations. In the example given the central position couples to its six neighbors, and the first star of one such neighbor is highlighted. This corresponds to a truncation of $n_{\text{max}} = 2$.}\label{fig:bz}
\end{figure}

Here we will outline the details of the Hamiltonian used to describe the graphene superlattice in the continuum limit, the electronic structure of which has been extensively studied \cite{MK14,WPMGF13}. 
In the particle-hole basis (which we distinguish from the eigenstates of the gapped Dirac Hamiltonian that are also commonly used in this field), the Dirac Hamiltonian for charge carriers of momentum $\vec{g}$ with Fermi velocity $v_F$ is given by
\begin{equation}
\mathcal{H}_D(\vec{g}) = \hbar v_F \vec{g}\cdot\vec{\sigma}.
\end{equation}
We decompose the momentum $\vec{g}$ within the graphene first Brillouin zone (FBZ) into a momentum $\vec{k}$ within the boundaries of the superlattice FBZ and a contribution from repeats of the superlattice FBZ, as
\begin{equation}
\vec{g}_n = \vec{k} + \vec{n}\cdot (\vec{G}_1, \vec{G}_2) \equiv \vec{k} + \vec{G}_n.
\end{equation}
Here $\vec{G}_n$ are the superlattice basis vectors of the  $n = 1 \ldots 6$ nearest neighbors in reciprocal space; these generate
the first harmonic functions of the superlattice \cite{WPMGF13}.
These are referred to 
as `one star of reciprocal lattice vectors' in the main text. Each star of 6 or 12 points in the reciprocal lattice consists of $\vec{G}$ vectors equivalent by symmetry. Successive stars are further apart from the $\vec{G} = 0$ origin. In this work we consider coupling up to $n_{\text{max}} = 5$  first stars: each position in our superlattice FBZ couples to its nearest neighbors via $\vec{G}_n$, which in turn couple to their nearest neighbors  for $n_{\text{max}}$ iterations. This is illustrated for $n_{\text{max}} = 2$ in Fig. \ref{fig:bz}. All calculations reported here are converged with respect to $n_{\text{max}}$ for the energy range shown; in this sense our analysis is \textit{non-}perturbative.

As outlined in the main text the effect of the graphene superlattice is treated as a perturbation on the Dirac Hamiltonian, such that the continuum limit is valid. This requires that the energies involved should be small when compared with the $\pi$ bandwidth, greater than the hopping matrix $t$ so that they lie in the energy range where the Dirac cone is defined. This is the case in the superlattices studied, where the
energy scales of typical perturbations $\sim$100 meV. The superlattice potential must also change change slowly, so that corrugations at scales comparable to the lattice constant of graphene, $a$ can be neglected. This assumption is satisfied in Moir\'e superlattices where the superlattice constant, $L$, is much larger than the lattice constant of graphene. This is the case for the superlattices studied in our paper, where L/a $\simeq$ 50. The superlattice potential, neglecting intervalley scattering, consists of seven terms \cite{WPMGF13}: a constant gap ($\delta$) and six spatially modulated potentials, including two scalar potentials ($V_s$), two mass gaps ($V_{\Delta}$) and two gauge field potentials ($V_g$), which each can be distinguished by their even ($e$) or odd ($o$) parity. Scalar and mass terms shift charge carrier energies at lattice sites dependent on their position in the superlattice; they have an either equal or opposite effect on each sublattice respectively. In the continuum limit, gauge fields describe modifications in the hopping energies due to the changing positions of lattice sites in the graphene layer. These discrete, real space deformations are represented by a gauge field and subsequent change of phase in the continuum. The resulting deformed lattice is shown in Fig. 1 of the main text; we shall further discuss the underlying vectors that produce this deformation in Section \ref{sec:nanoribbon}. In our notation, neighboring sites $n$ and $n'$ couple via the first star of reciprocal lattice vectors by
\begin{gather}
\mathcal{H}_\text{pert} = \sum_{j=1}^6 V_{\vec{G}_j} \delta_{\vec{g}_n - \vec{g}_{n'},\vec{G}_j}, \\
V_{\vec{G}_j} = \big( V_s^e + i (-1)^j V_s^o\big) \mathbb{I}_2 + \big( V_{\Delta}^o + i (-1)^j V_{\Delta}^e \big) \sigma_3 + \big(i V_g^e + (-1)^j V_g^o \big) \left( \begin{matrix} 0 & -ie^{-i\phi_{\vec{G}_j}} \\ ie^{i\phi_{\vec{G}_j}} & 0 \end{matrix} \right),\label{eq:superlattice}
\end{gather}
where the additional phase $\phi_{\vec{G}_j} = \arg \left({G}_{j,x}+i{G}_{j,y}\right)$. This allows us to write the overall Hamiltonian as the original Dirac Hamiltonian plus a small correction due to the superlattice,
\begin{align}
\begin{split}
\mathcal{H} & = \mathcal{H}_D(\vec{g}_n)\otimes \mathbb{I}_N + \mathcal{H}_\text{pert} \\
& = \mathcal{H}_D(\vec{k})\otimes \mathbb{I}_N + \mathcal{H}_D(\vec{G}_n)\otimes \mathbb{I}_N + \mathcal{H}_\text{pert} \\
& = \mathcal{H}_D(\vec{k})\otimes \mathbb{I}_N + \mathcal{H}_{SL}.
\end{split}
\end{align}
Results in the main text use superlattice parameters as derived in \cite{SGSG14}: for a strained graphene superlattice with an associated vector field, we take
\begin{equation}\label{eq:parameters}
(V_s^e, V_s^o, V_{\Delta}^e, V_{\Delta}^o, V_g^e, V_g^o, \delta) = (21, 38, 6, 0, -42, -21, 50)\,\mathrm{meV}.
\end{equation}
We have verified that results and conclusions are qualitatively unchanged for any generic combination of superlattice potentials and that flat bands can be produced using all superlattice parameters, with the exception of the gauge potentials $V_g$ that do not influence edge modes. Gauge potentials contribute an additional phase, which can be removed for any well-localized state by an appropriate 
change-of-gauge transformation; thus $V_g$ will not significantly alter the flat bands found in our tight-binding calculations.

\section{Chern Numbers\label{sec:Chern}}

To calculate the Chern numbers of each subband we use the method outlined by Fukui \emph{et al} \cite{FHF05}. The superlattice Brillouin zone is tessellated with hexagonal plaquettes, each of which contributes to the Berry curvature. The Berry curvature from each plaquette $p$ 
depends on its wavefunction $\psi_p$, in particular the product of the wavefunction overlaps between neighboring plaquettes 
$\bra{\psi_{p+1}}\ket{\psi_p}$: Berry curvature is determined up to a factor of 2$\pi$ by the argument of this complex product. 
Rescaling by $2\pi$ gives the Chern number contribution from this plaquette, the sum of 
which over the Brillouin zone gives the Chern number for a sub-band.

Plaquettes at the edges of the Brillouin zone are dealt with separately, as we cannot perfectly tessellate hexagonal plaquettes within the boundaries of the hexagonal Brillouin zone. Due to the periodicity of the Brillouin zone, plaquettes at the edges contribute a fraction of their Berry curvature determined by the area of the plaquette within the Brillouin zone. Plaquettes at the side and corners of the Brillouin zone contribute one half and one third of their Berry curvatures respectively.

Numerical errors may produce Chern numbers that are only approximately integers: the typical numerical accuracy of the Chern numbers depends on $n_{\text{max}}$, but we find machine accuracy integers for $n_{\text{max}}$ as small as $5$.

\begin{figure}
	\begin{tabular}{ccl}
	\Large Coarse & \Large Fine \\
	\includegraphics[width=0.3\textwidth,valign=c]{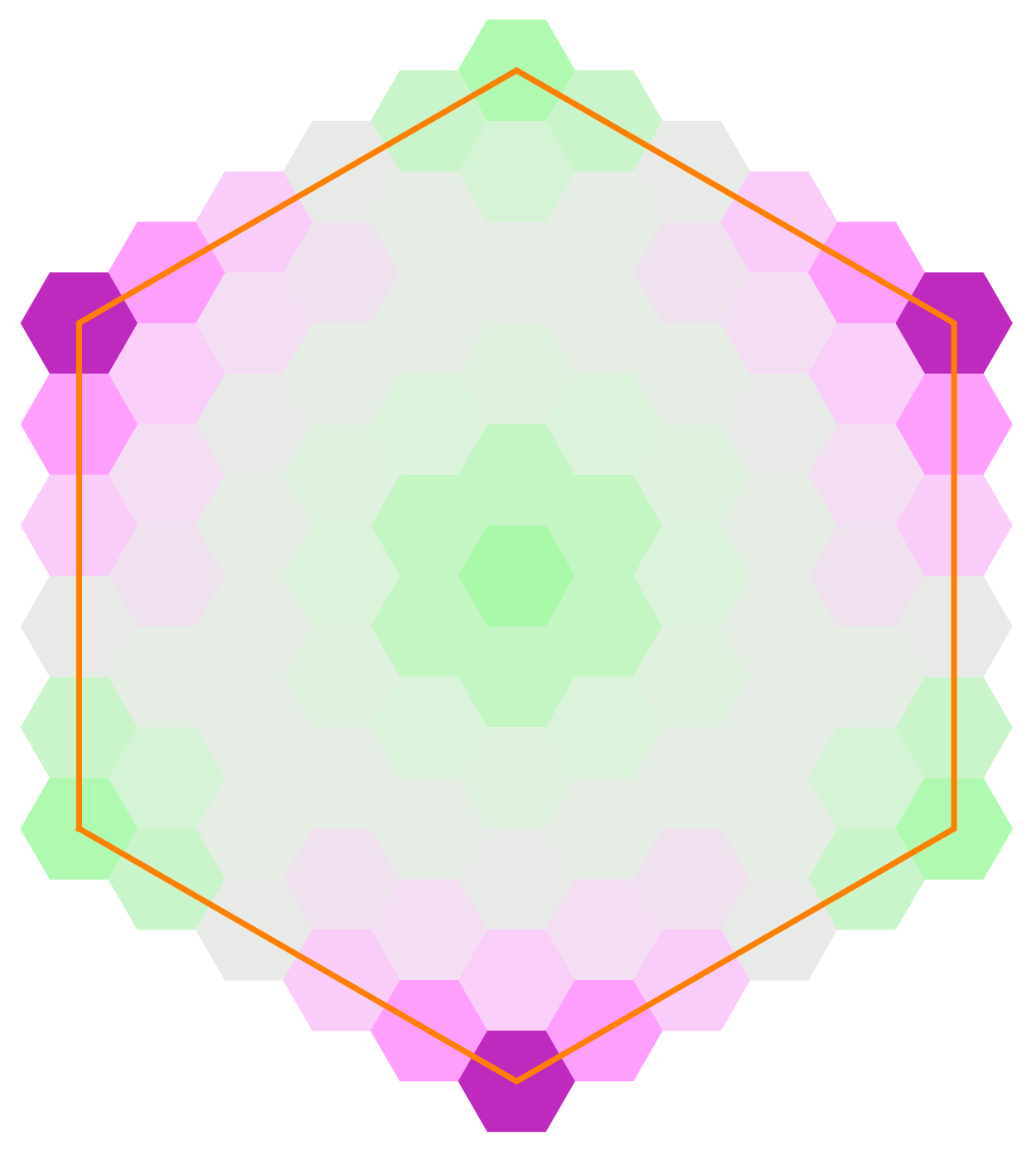}&
	\includegraphics[width=0.3\textwidth,valign=c]{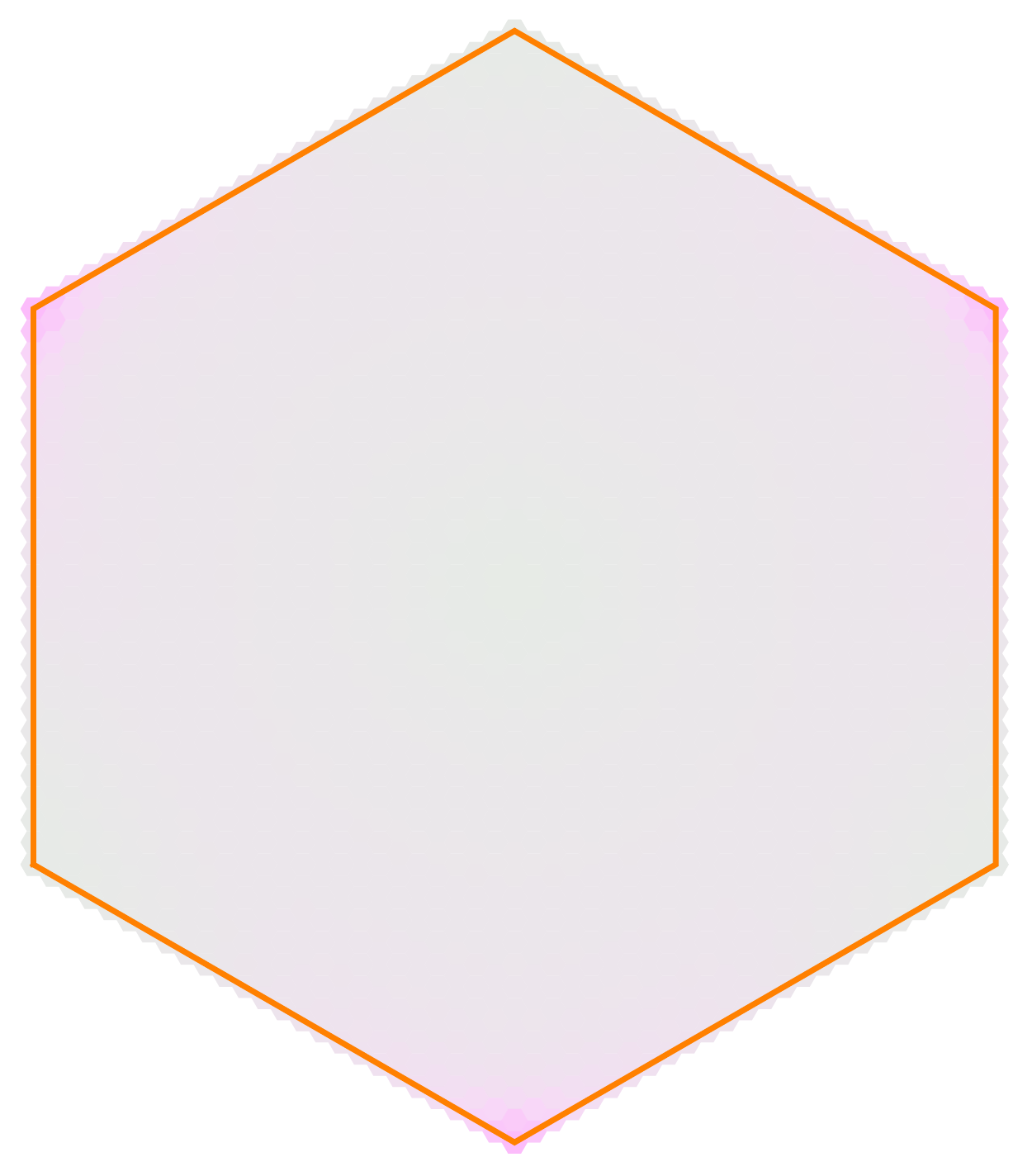}	\\
	\includegraphics[width=0.3\textwidth,valign=c]{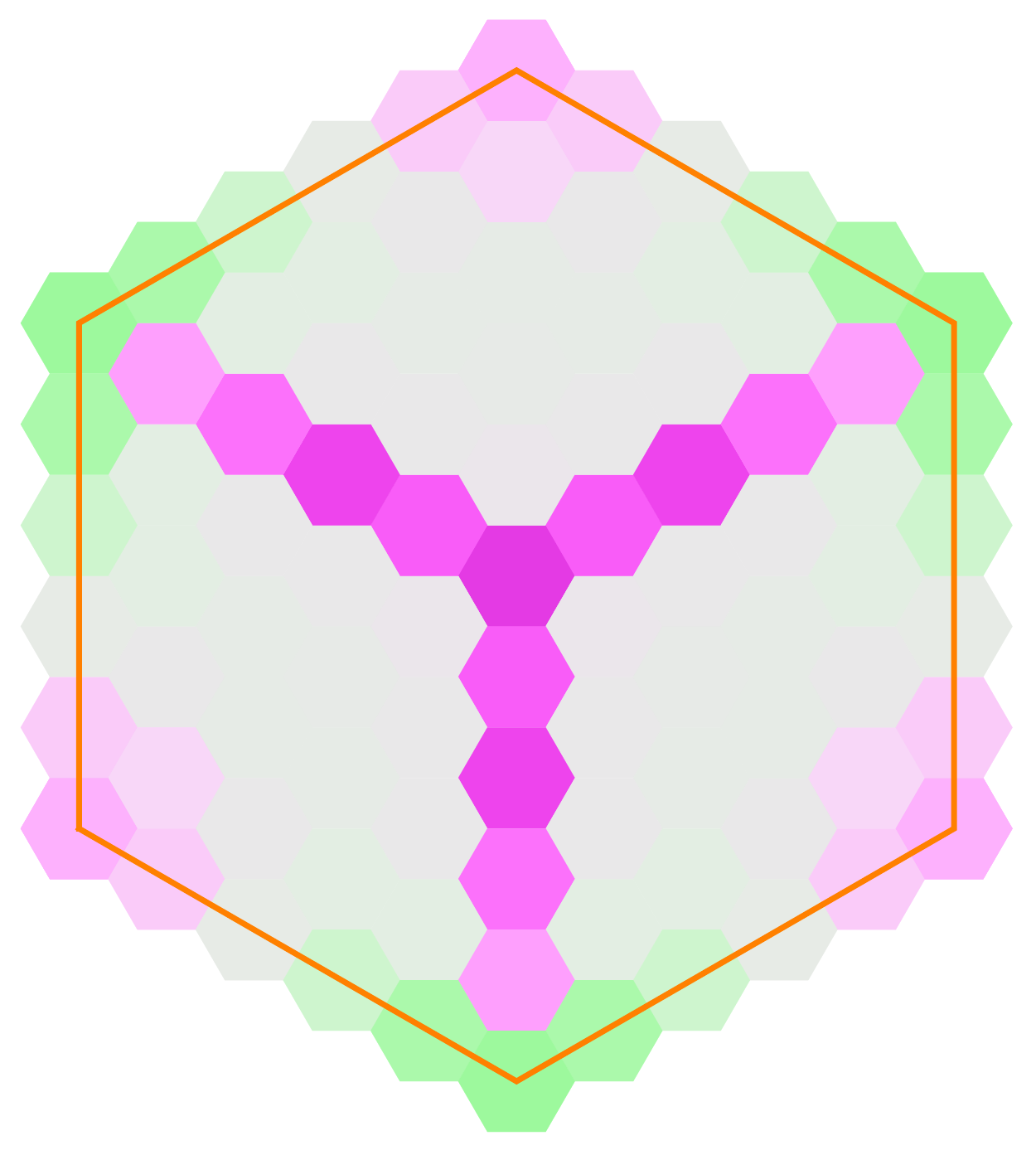}&
	\includegraphics[width=0.3\textwidth,valign=c]{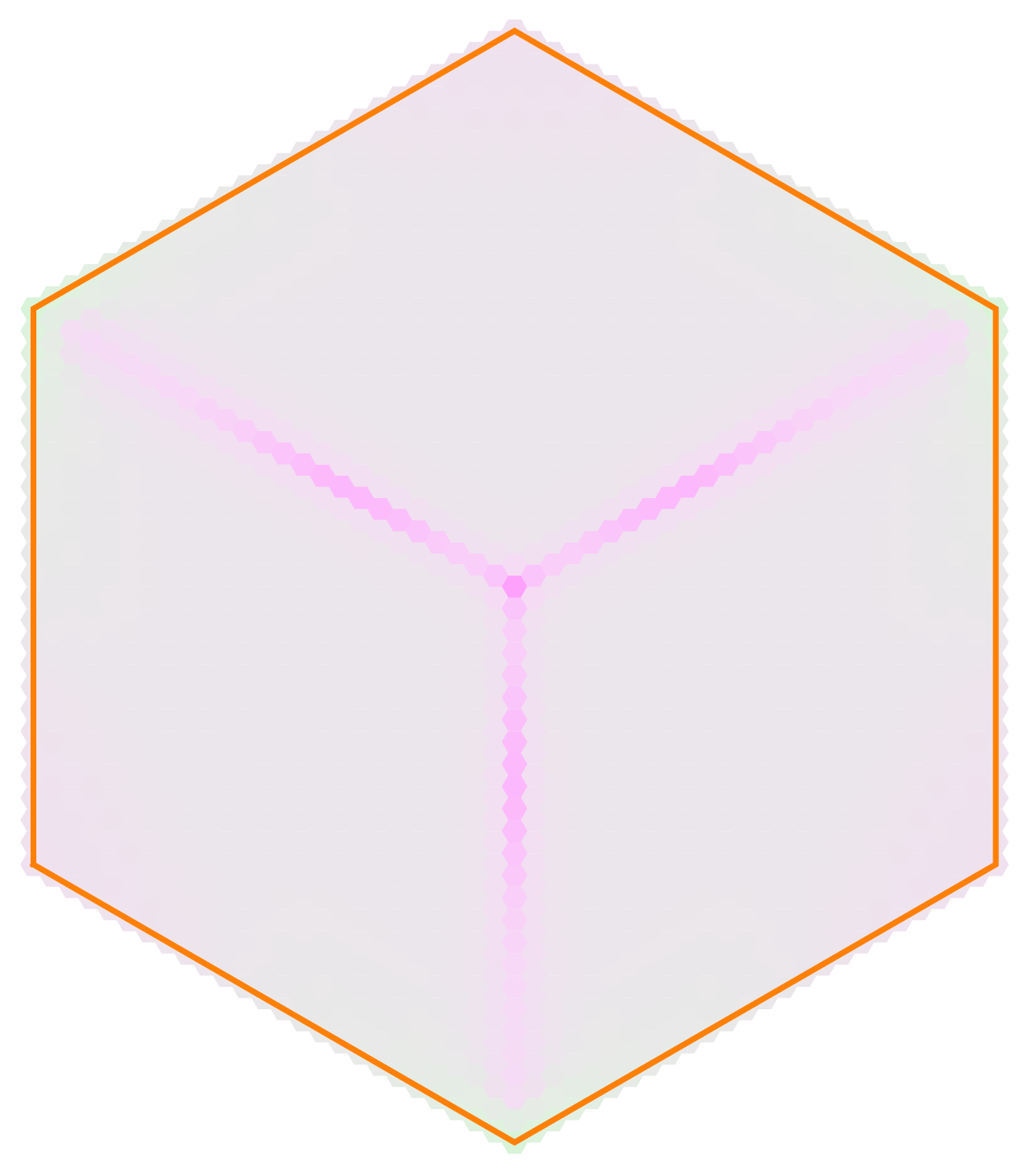}& \\
	\includegraphics[width=0.3\textwidth,valign=c]{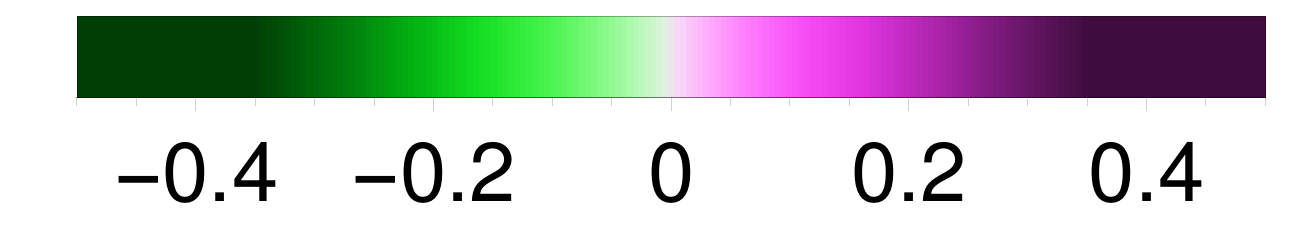}&
	\includegraphics[width=0.3\textwidth,valign=c]{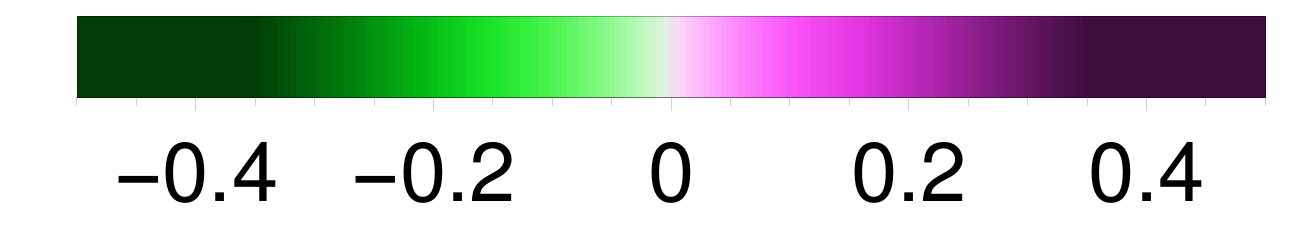}
	\end{tabular}
	\caption{Berry curvatures for the conduction (\textit{left}) and first excited (\textit{right}) subbands, using both 10 and 50 plaquettes to span the width of the Brillouin zone (a total of 75 and 1875 total plaquettes respectively). The orange border marks the edge of the Brillouin zone, beyond which contributions are not included. The Chern numbers for the subbands are 0 and 1 for either choice of hexagonal grid.} \label{fig:chern}
\end{figure}

Even though the answers are thus always integers, a potential issue with this technique is caused by the coarseness of the tessellation of the Brillouin zone, which can lead to the \emph{wrong} integer value! Sufficiently large plaquettes can result in a phase change over an individual plaquette larger than $2\pi$. This which will produce an integer change in the Chern number when compared to the same calculation over a finer grid of hexagons, which will include smaller phase changes per plaquette. We have verified that our calculations are robust against this problem of topological charge `falling through the lattice' by taking increasingly finer plaquettes and showing that the Chern numbers remain unchanged. Example results for the conduction and first excited subbands using superlattice parameters from Eq. \ref{eq:parameters} are given in Fig.~\ref{fig:chern} for coarse and fine grids across the Brillouin zone.

\newpage
\section{Superlattice Nanoribbons}\label{sec:nanoribbon}

\begin{figure}
	\includegraphics[trim={0 0.0355cm 0.012cm 0},clip,width=0.2\textwidth]{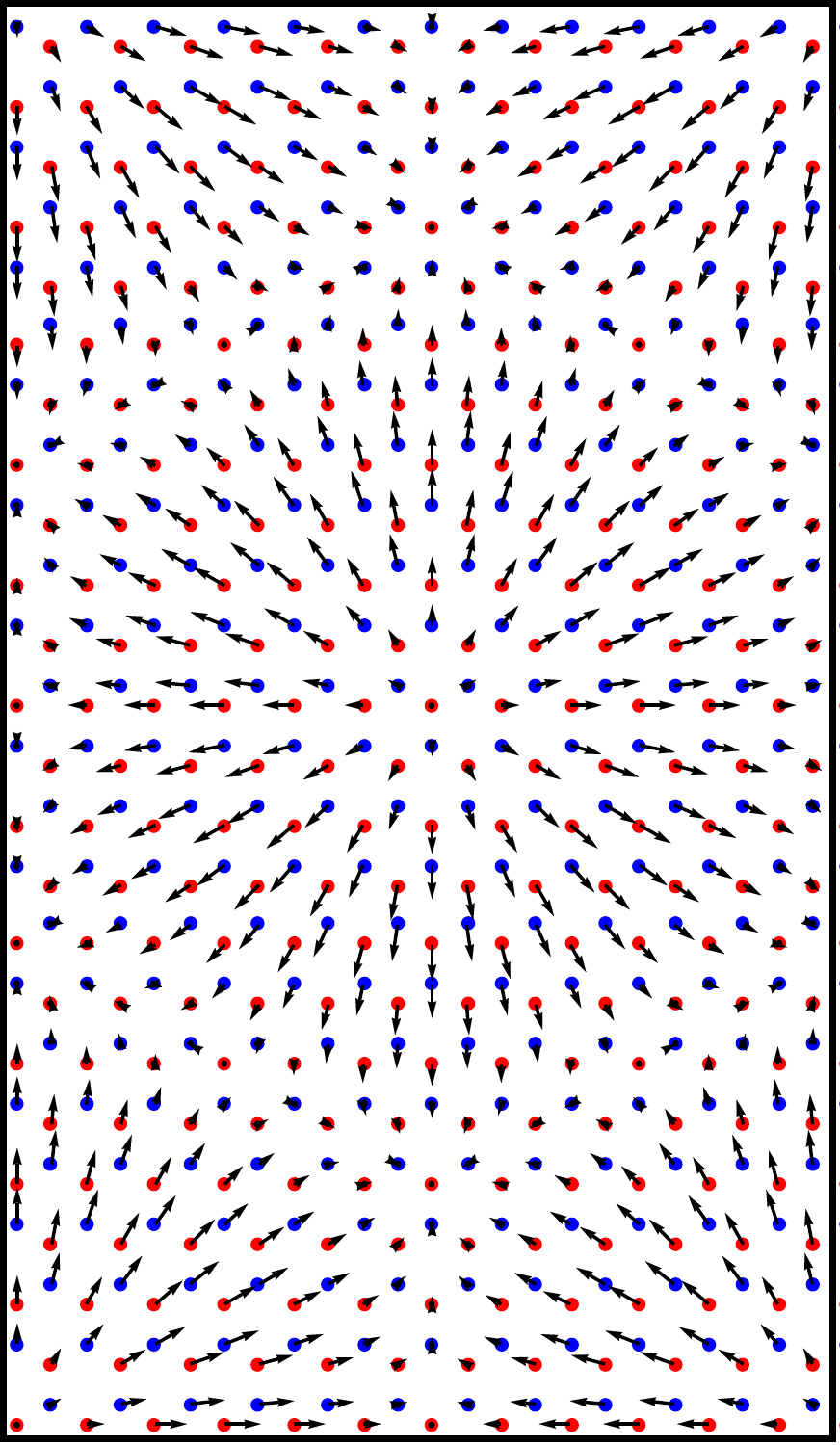}
	\includegraphics[trim={0 0.0355cm 0.011cm 0},clip,width=0.2\textwidth]{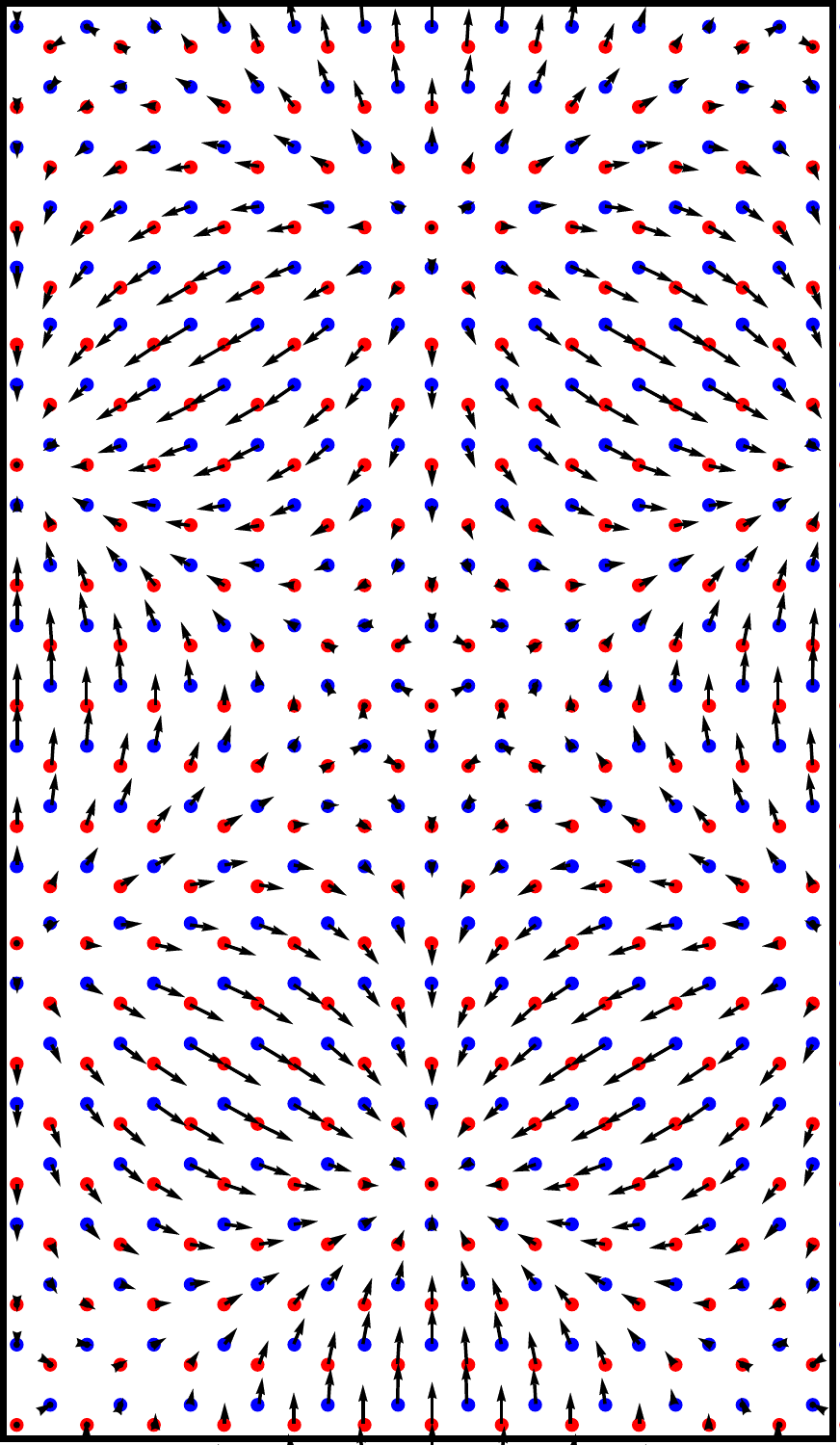}
	\caption{The even (\textit{left}) and odd (\textit{right}) gauge potentials corresponding to a superlattice with a unit cell $L = 12a$, where $a$ is the graphene unit cell. Vectors indicate the strength of the gauge field at each lattice site, caused by the superlattice displacing individual sites and locally deforming bonds. In either case the corresponding potential has strength 0.5 eV, while A (B) sublattice sites are shown in red (blue).}\label{fig:vec}
\end{figure}

We can also investigate superlattice effects using a tight-binding model by including the discrete analogues of the superlattice potentials from Eq.~\ref{eq:superlattice}. These are modeled by a modulation of the hopping parameters in the tight-binding model, due to the superlattice modulation of the carbon-atom positions. The spatially modulated superlattice potentials may identically or oppositely perturb the A and B sublattices, in the case of the scalar potentials and mass gaps respectively. The gauge field potentials used in the continuum model are caused by a change in the local bond lengths in the bond direction $e_{ij}$. This induced strain can be described in the tight-binding model by altering the hopping parameter. The superlattice parameters used in the continuum model are directly related to those in the tight-binding model presented here: we assume that the continuum parameters directly map to those of our tight-banding model and can be applied without alteration, following the good agreement between cases found in \cite{Weckbecker16}. In this formulation the Hamiltonian can thus be expressed as 

\begin{equation}\label{eq:tb}
\mathcal{H} = - (t+V_g t) \sum_{\langle i,j \rangle } (a_{i}^\dagger b_{j} + b_{i}^\dagger a_{j}) + V_s + V_{\Delta} + \delta,
\end{equation}
where $a_{i}, a_{i}^\dagger (b_{i}, b_{i}^\dagger)$ are the creation and annihilation operators for electrons on sublattice A (B) at site $\vec{r}_i$ and graphene's nearest-neighbor hopping parameter $t=2.74$ eV \cite{KR09}. The full form of the superlattice potentials is
\begin{gather}
V_s = V_s^e \sum_{l=1}^3 \cos(\vec{g_l}\cdot \vec{r}_j) + V_s^o \sum_{l=1}^3 \sin(\vec{g_l}\cdot \vec{r}_j),
\\ V_{\Delta} = \pm \left( V_{\Delta}^e \sum_{l=1}^3 \sin(\vec{g_l}\cdot \vec{r}_j) + V_{\Delta}^o \sum_{l=1}^3 \cos(\vec{g_l}\cdot \vec{r}_j) \right),
\\ V_g = V_g^s \sum_{l=1}^3 \vec{g_l}\left( \sin(\vec{g_l}\cdot \vec{r}_i) - \sin(\vec{g_l}\cdot \vec{r}_{j})\right)e_{ij} + V_g^o \sum_{l=1}^3 \vec{g_l}\left(\cos(\vec{g_l}\cdot \vec{r}_i) - \cos(\vec{g_l}\cdot \vec{r}_{j})\right)e_{ij},
\end{gather}
where $\vec{g_l}$ are the reciprocal lattice vectors of the triangular superlattice which determine the modulation strength at each lattice position; the even and odd vector contributions for the example of a small superlattice are given in Fig. \ref{fig:vec}. $V_{\Delta}$ has the opposite sign for each sublattice.

\begin{figure}
	\includegraphics[width=0.4\textwidth]{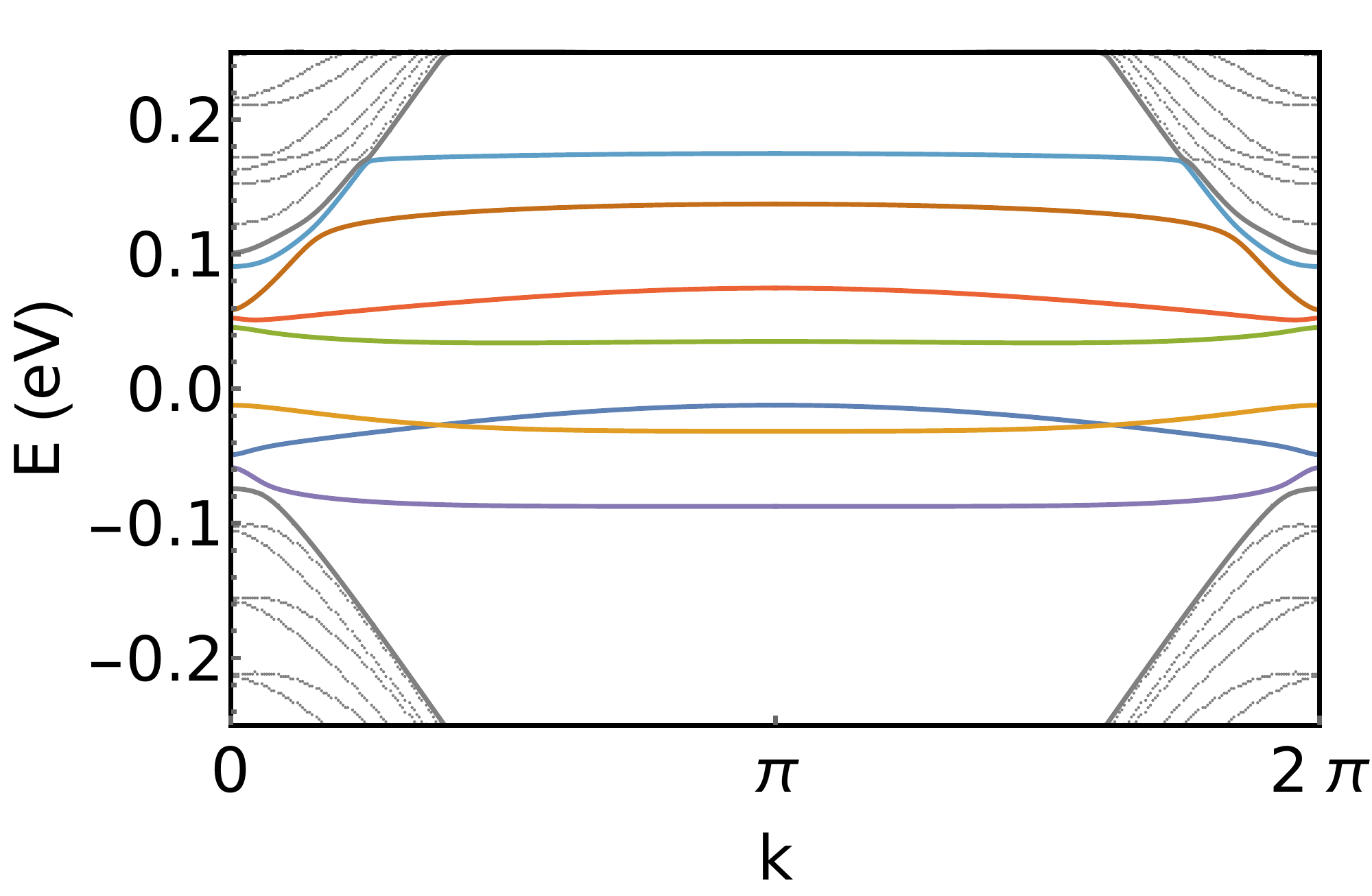}
	\includegraphics[width=0.4\textwidth]{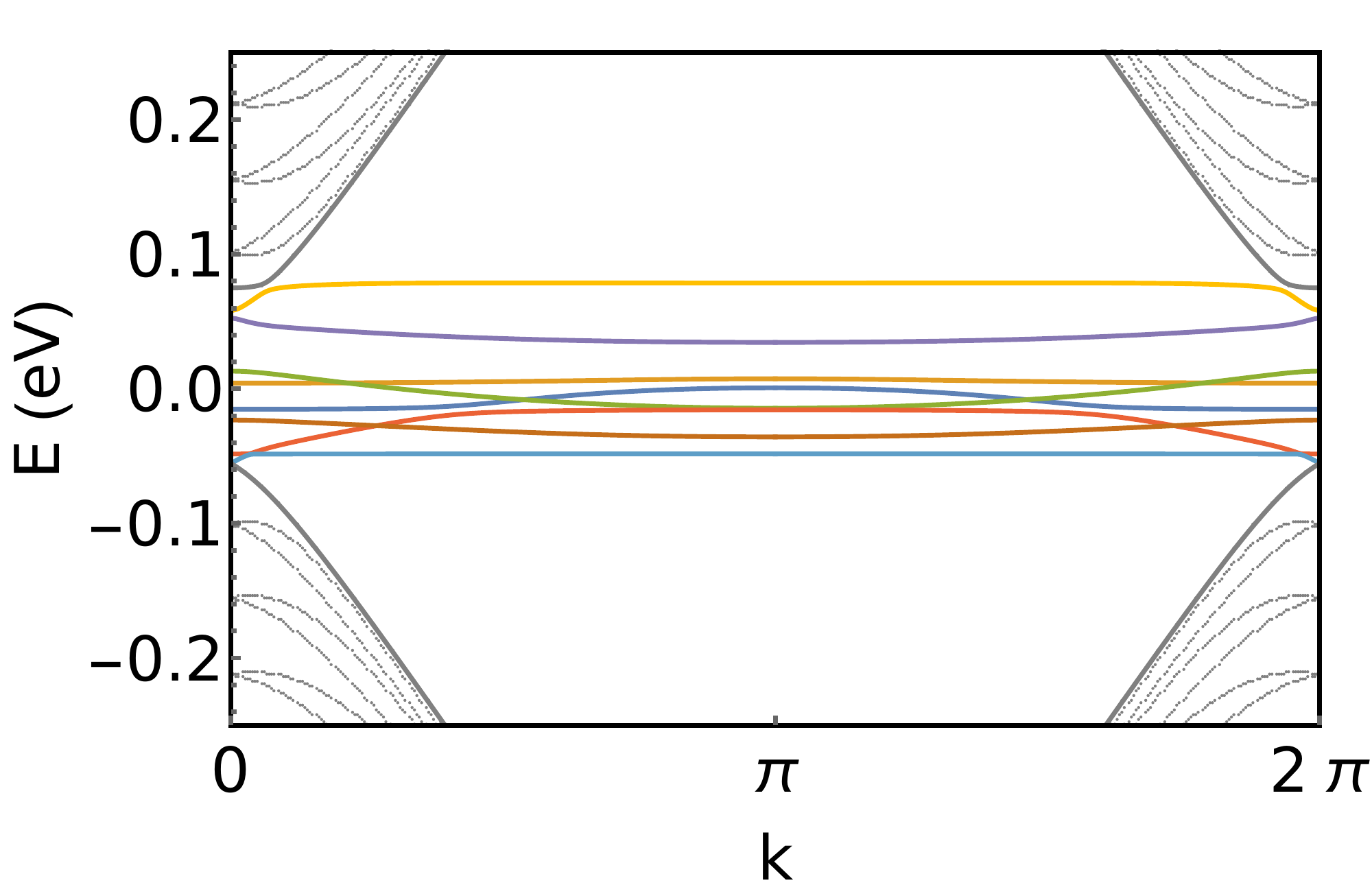}
	\includegraphics[width=0.4\textwidth]{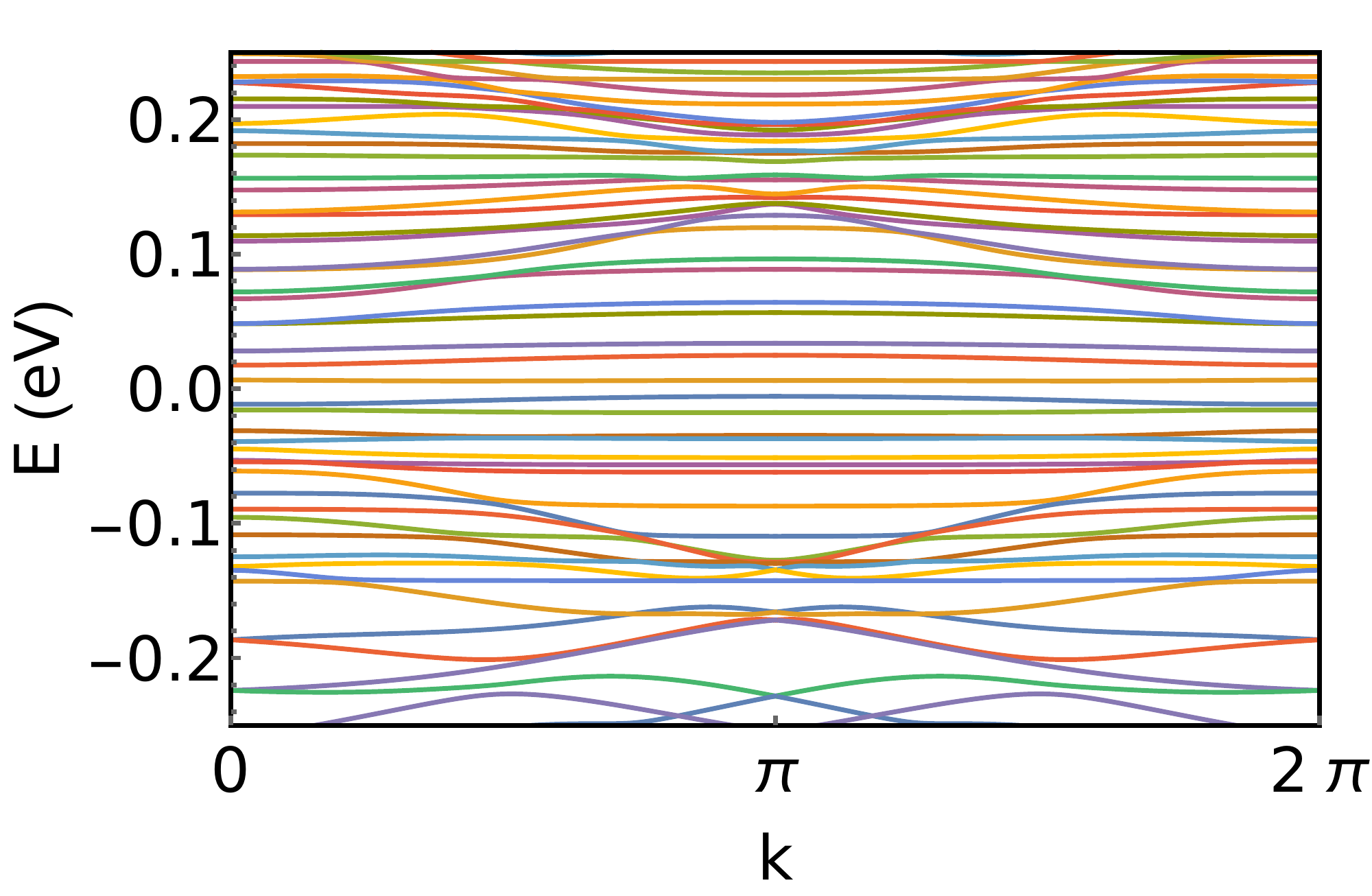}
	\includegraphics[width=0.4\textwidth]{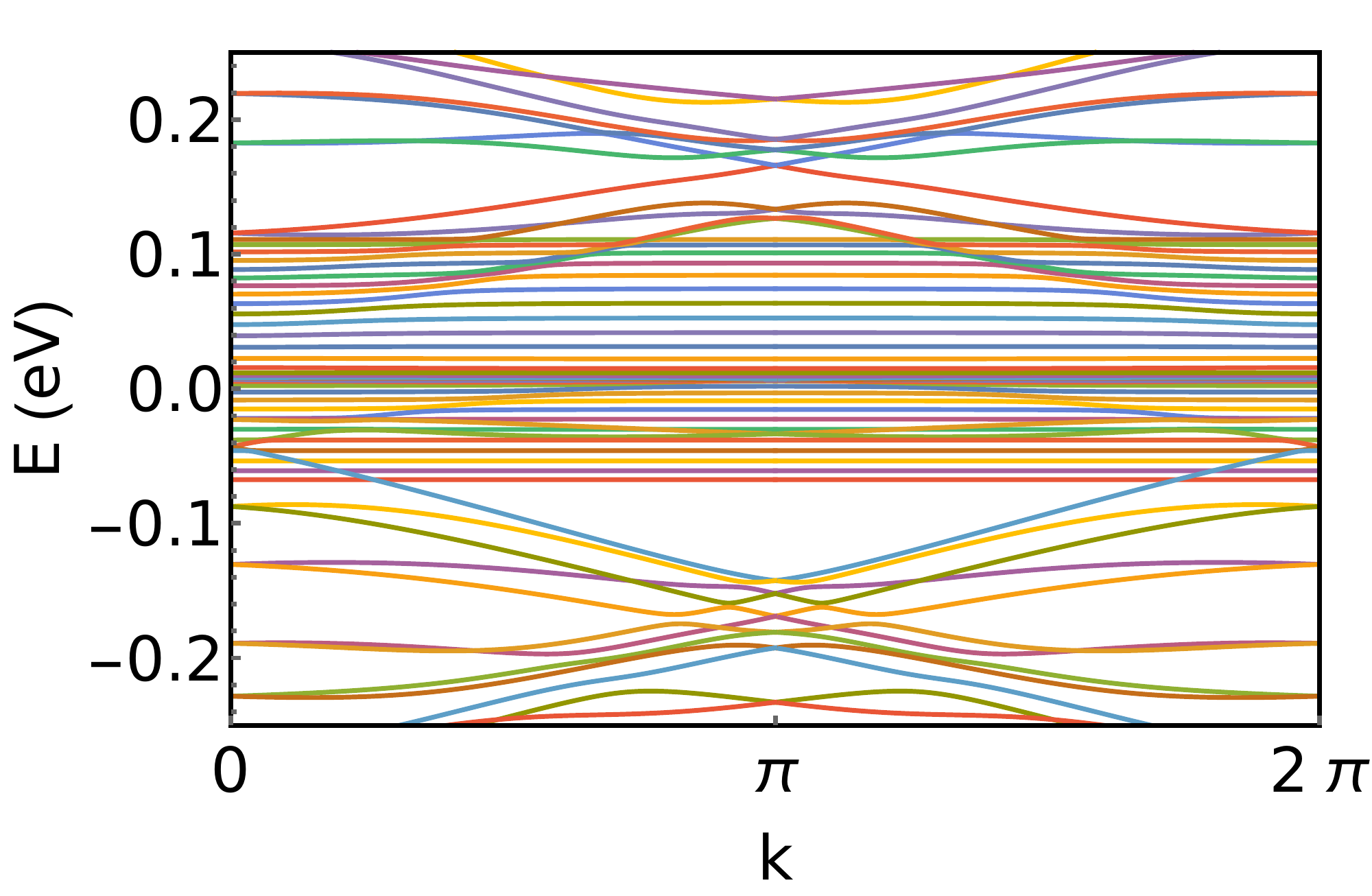}
	\caption{Band structures for a nanoribbon 144 lattice sites in width, the superlattice unit cell size 12 $\times$ 12 (\textit{top}) and 48 $\times$ 48 (\textit{bottom}) graphene unit cells. Superlattice parameters
used include a strong even (\textit{left}) and odd (\textit{right}) scalar potentials: here  $(V_s^e,V_s^o)$ = (100, 38) meV and $(V_s^e,V_s^o)$ = (21, 100) meV in each case respectively, while the other parameters remain as in Eq. \ref{eq:parameters}.}\label{fig:3x48}
\end{figure}

To analyze this tight-binding model we construct a semi-infinite nanoribbon with periodic boundaries in the direction of the axis parallel to the superlattice. The one-dimensional Brillouin zone is parallel to the nanoribbon direction: we solve Eq. \ref{eq:tb} for charge carriers with momentum $0 \leq k \leq 2\pi$, where $k = k_{\parallel}d$ for momentum $k_{\parallel}$ in the direction of a nanoribbon of unit cell $d$. Points $K$ and $K'$ in the continuum Brillouin zone lie at $k=0$ here. Results in the main text use nanoribbons 144 lattice sites in width with zigzag edges and with superlattice unit cells of 48 $\times$ 48 graphene unit cells, using the superlattice parameters in Eq. \ref{eq:parameters}. Additional band structures for stronger scalar potentials ($V_s^{e,o} =  0.1\,\mathrm{eV}$ with other parameters unchanged) are given in Fig.~\ref{fig:3x48}. The appearance of flat bands in both these and other cases indicates that edge transport is a generic feature of nanoribbons with zigzag edge configurations, irrespective of the specific parametrization of the superlattice. This has been verified for many other ``generic" superlattice potentials, as we outline in the following section.

\begin{figure}
	\includegraphics[width=0.4\textwidth]{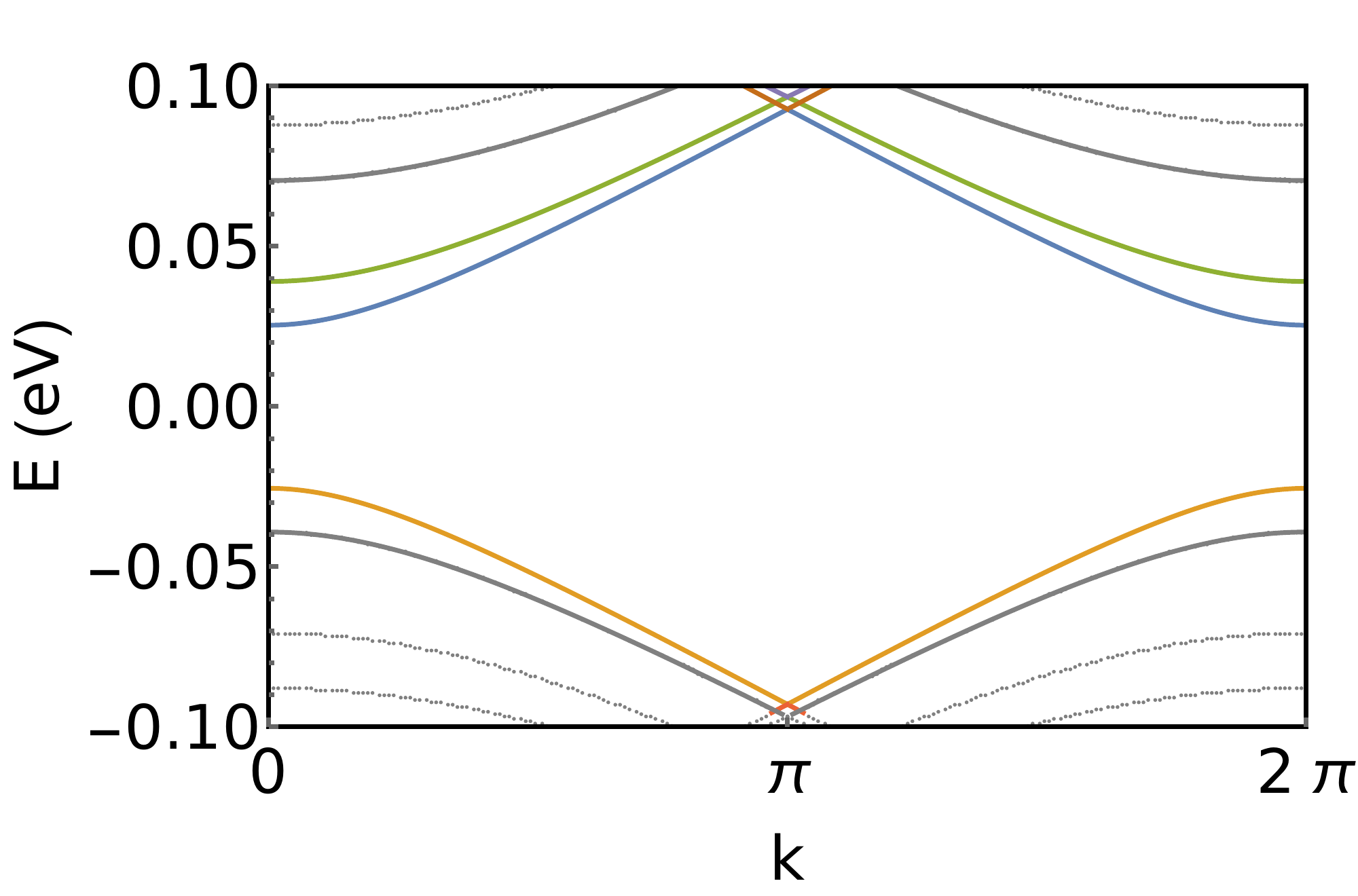}
	\caption{The band structure for a graphene nanoribbon with armchair edges and superlattice parameters from Eq. \ref{eq:parameters}. The nanoribbon width and superlattice unit cell sizes are 144 and 48 $\times$ 48 graphene unit cells respectively.}\label{fig:armchair}
\end{figure}

We have repeated these calculations for nanoribbons with armchair edge configurations, demonstrating that the superlattice does not create new edge modes in this system and that the gap persists. Results for a realistically large system with a 48 $\times$ 48 supercell are given in Fig.~\ref{fig:armchair}.

\newpage
\section{Density of Midgap States}

\begin{figure}
	\includegraphics[width=0.3\textwidth]{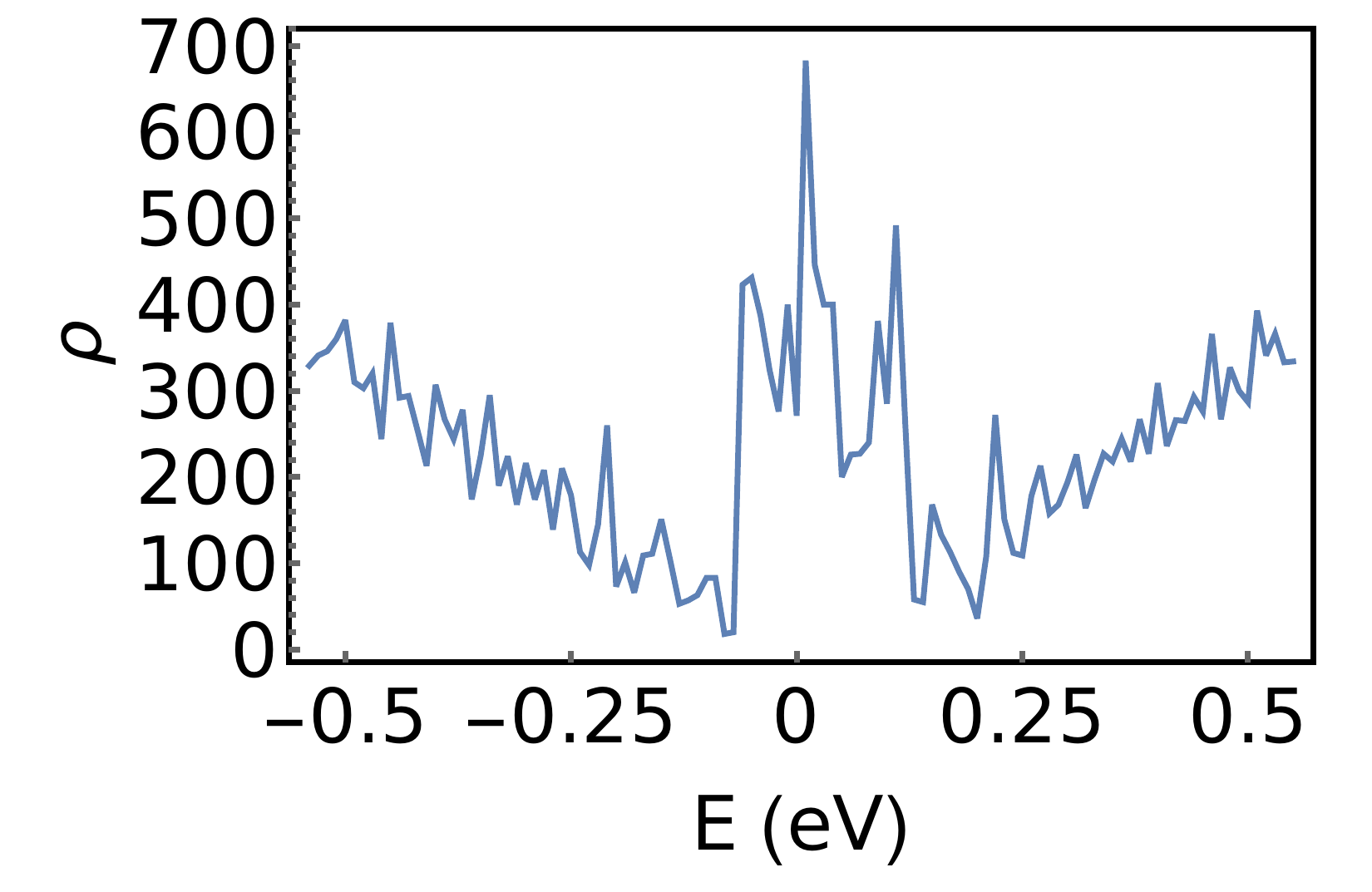}
	\includegraphics[width=0.3\textwidth]{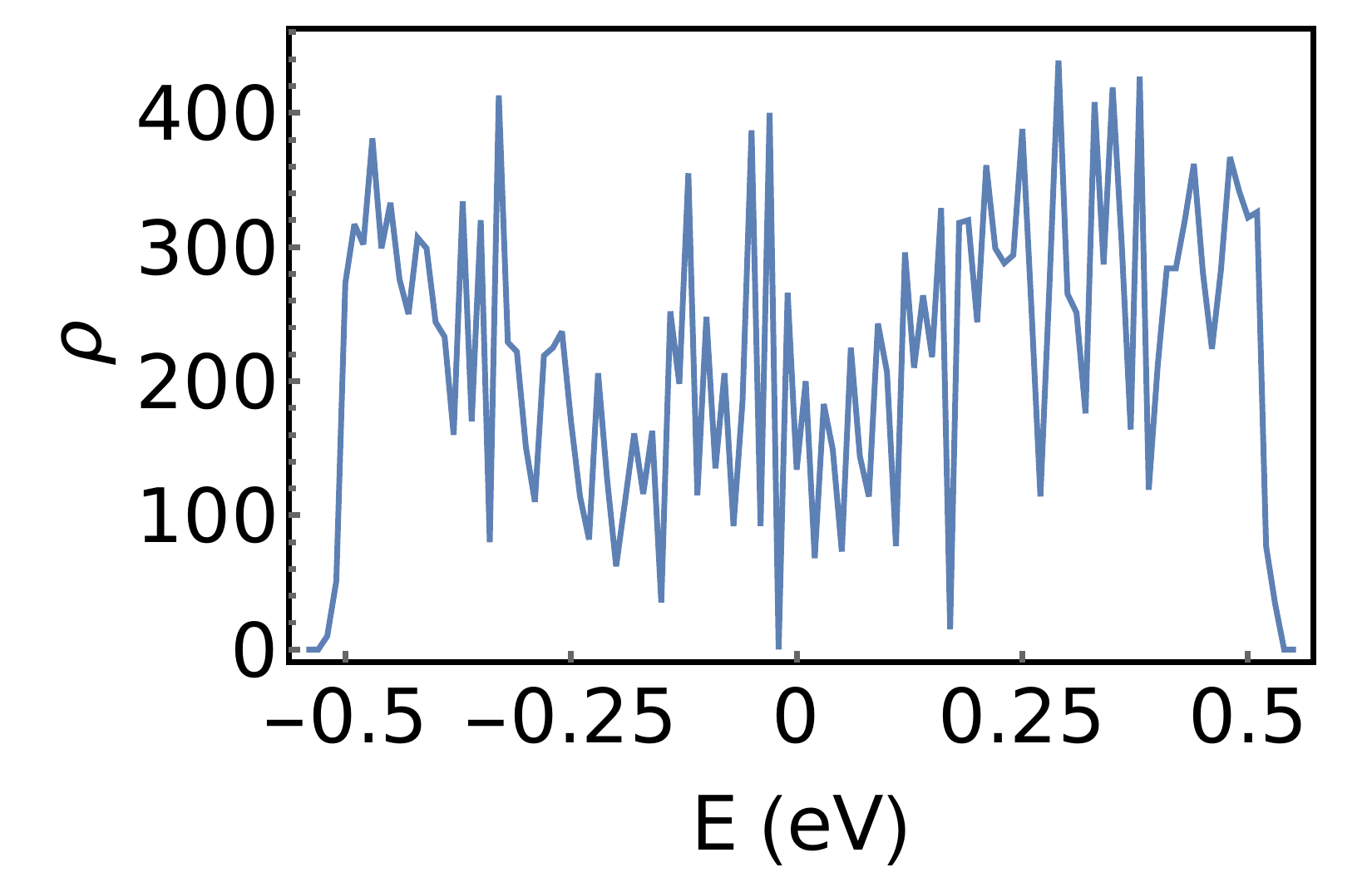}
	\includegraphics[width=0.3\textwidth]{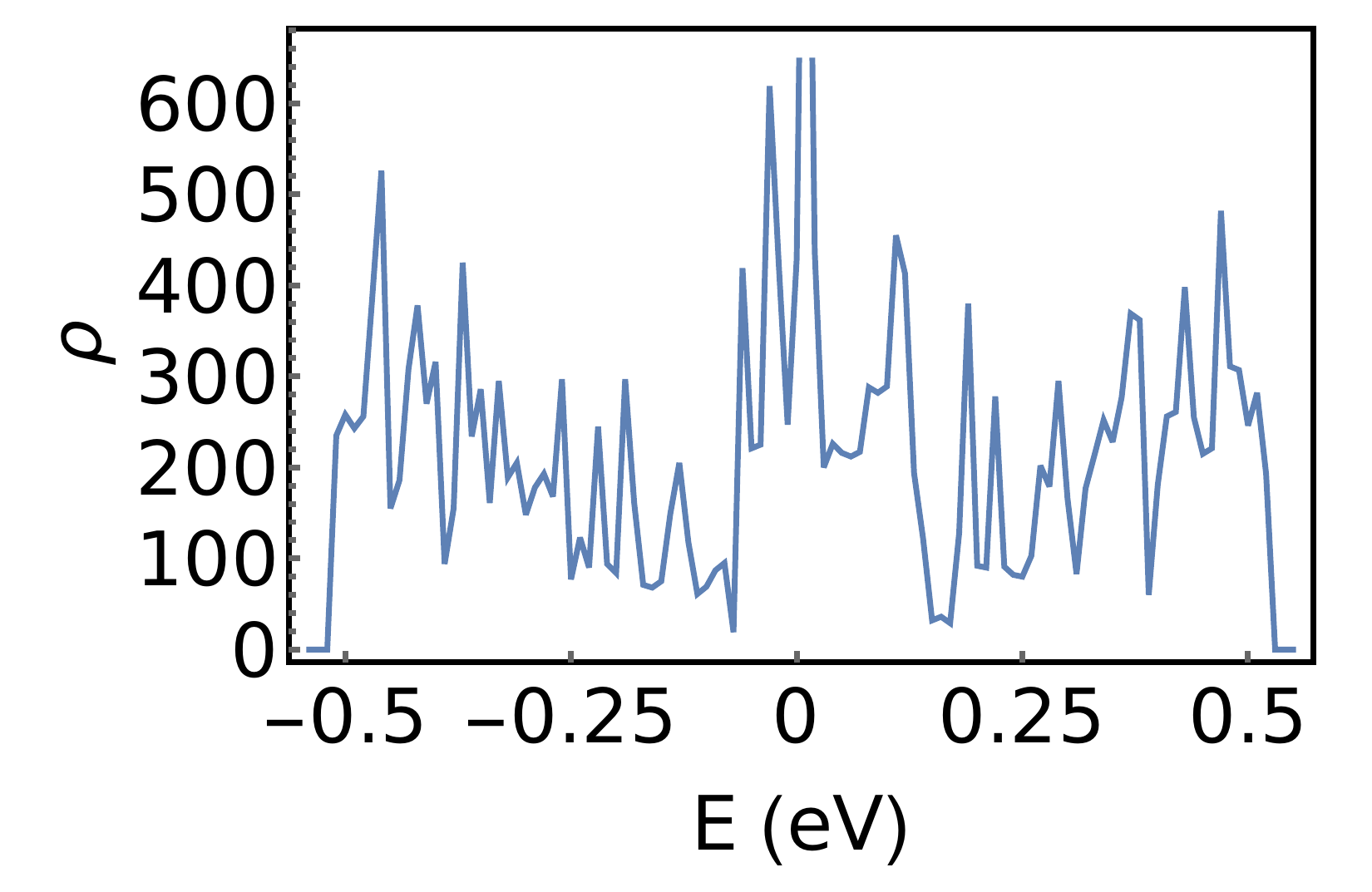}
	\caption{The density of states for nanoribbons 144 lattice sites in width and superlattice unit cells of 48 $\times$ 48 graphene cells. Superlattice parameters are those given in Eq. \ref{eq:parameters} (\textit{left}), including cases with large even (\textit{middle}) and odd (\textit{right}) scalar potentials, $V_s^{e,o} = 100$ meV. Each density of states consists of 28\%, 15\% and 28\% edge modes respectively.}
\label{fig:dos}
\end{figure}

Edge states in superlattices with realistically large unit cells occupy a significant fraction of the bulk gap, as we can see in Fig. \ref{fig:3x48}. We estimate the fraction of all states involved in edge transport by calculating the density of states, comparing the number of states in the gap region (here taken to be $-0.1 \leq E \leq 0.1$ eV for consistency with Fig. \ref{fig:3x48}, and Fig. 3 in the main text) to those over the entire energy range $-0.5 \leq E \leq 0.5$ eV. The density of states and percentage involved in edge transport for nanoribbons with large 48 $\times$ 48 graphene unit cell supercells are given in Fig. \ref{fig:dos}. We make use of the parameters given in Eq. \ref{eq:parameters}, as well as the parameter set with larger even/odd scalar potentials $V_s^{e,o} = 100$ meV as in Fig. \ref{fig:3x48}, and find that $15\% - 28\%$ of states are involved in edge transport.

Descriptions of realistic superlattices will involve a combination of all parameters in our model (Eq. \ref{eq:tb}). To demonstrate that generic superlattice perturbations will exhibit significant midgap transport, we calculate the fraction of midgap states for individual superlattice parameters $V_{s,\Delta}^{e,o}$ in the range $5 \leq V_{s,\Delta}^{e,o} \leq 20$ meV, with a gap of $100$ meV: results are given in Table \ref{tab:dos}. The exception is the vector potentials, which are not included here as they can be arbitrarily removed for edge states with an appropriate choice of gauge. We find that each case hosts a large number of edge modes, typically 30\% of those in the given energy range. Combinations of superlattice parameters tend to produce more modes inside the gap region: a given example of a small even scalar addition to an otherwise odd parity scalar potential increases the fraction of midgap states by $\sim10\%$, as shown in the table. From this we expect that many edge modes should be present in the previously gapped region for any general combination of superlattice parameters.

\begin{table}
	\caption{Fraction of the total density of states calculated in the range $-0.5 \leq E \leq 0.5$ eV that can be attributed to edge modes (taken to lie in the bulk gap, range $-0.1 \leq E \leq 0.1$ eV). All calculations are performed for nanoribbons 144 lattice sites in width, with superlattice unit cells of 48 $\times$ 48 graphene cells. Unless specified otherwise all superlattice parameters are set to $0$, while all cases have a bulk gap $\delta = 0.05 eV$. All superlattice parameter strengths are given in meV.}\label{tab:dos}
 \begin{center}
\begin{tabular}{ |c| c c c| c c c| c c c| c c c| c c c| }
 \hline {Parameter strength} & \multicolumn{3}{c|}{$V_s^e$} & \multicolumn{3}{c|}{$V_s^o$} & \multicolumn{3}{c|}{$V_\Delta^e$} & \multicolumn{3}{c|}{$V_\Delta^o$} & {$V_s^e$} & {$V_s^o$} & {$\delta$} \\
 {(meV)} & 5 & 10 & 20 & 5 & 10 & 20 & 5 & 10 & 20 & 5 & 10 & 20 & 8 & 20 & 20 \\ \hline 
 Midgap states (\%) & 30 & 30 & 27 & 30 & 30 & 30 & 30 & 30 & 30 & 30 & 30 & 25 & & 40 & \\  
 \hline	
\end{tabular} \end{center}
\end{table}

\section{Changing Supercell Size}

In order to address the most experimentally relevant superlattice configurations of commensurate graphene and hexagonal boron nitride, we have used large superlattice unit cells including approximately 50 $\times$ 50 graphene unit cells, while the angle between their crystallographic axes that determines the size of the Moir\'e pattern $\theta \sim 0^{\circ}$. By increasing $\theta$ the size of the superlattice unit cell is reduced, and we have verified that this effect does not significantly alter the conclusions drawn in the main text, as we shall outline here.

\begin{figure}
	\includegraphics[width=0.3\textwidth]{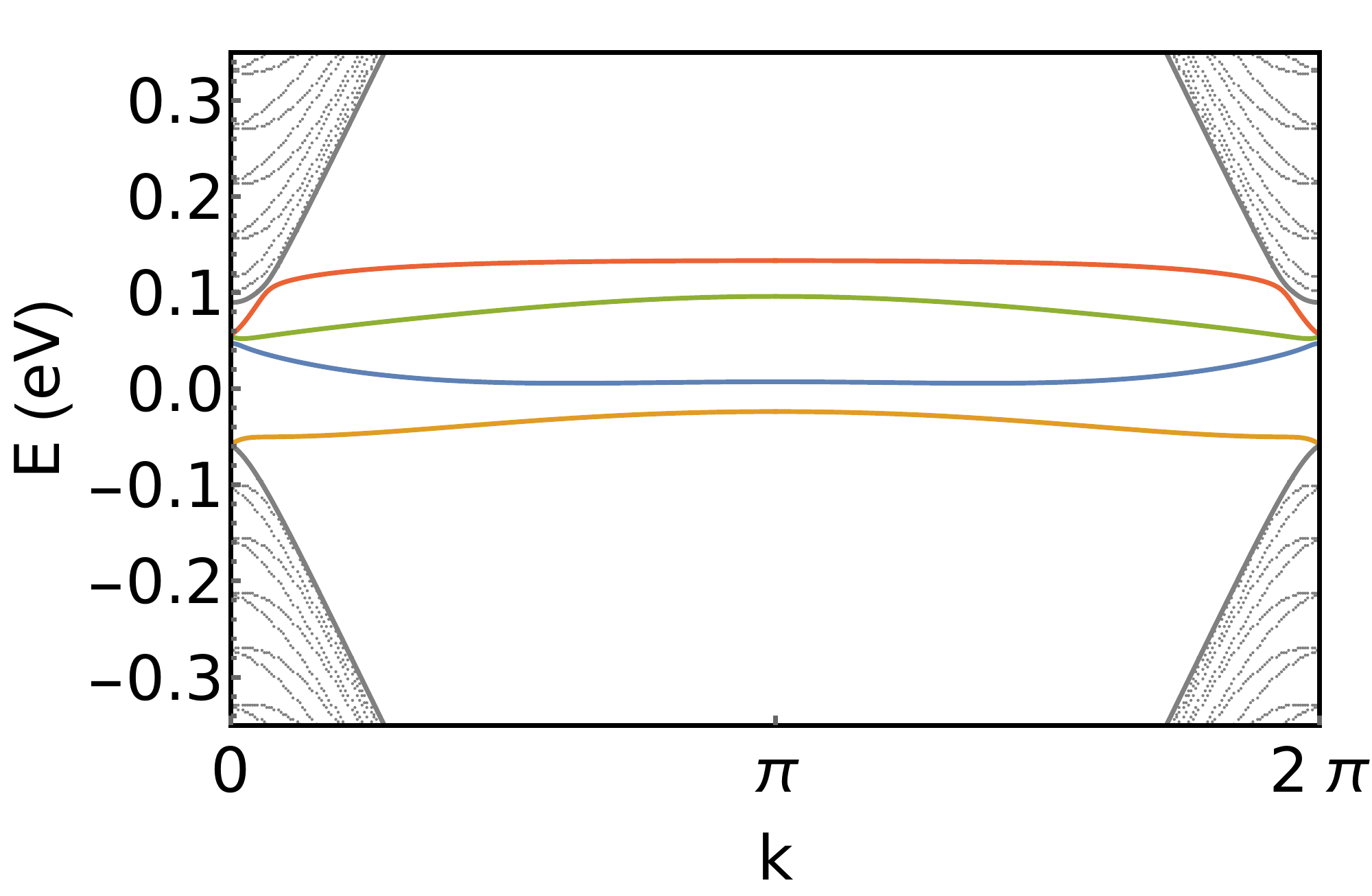}
	\includegraphics[width=0.3\textwidth]{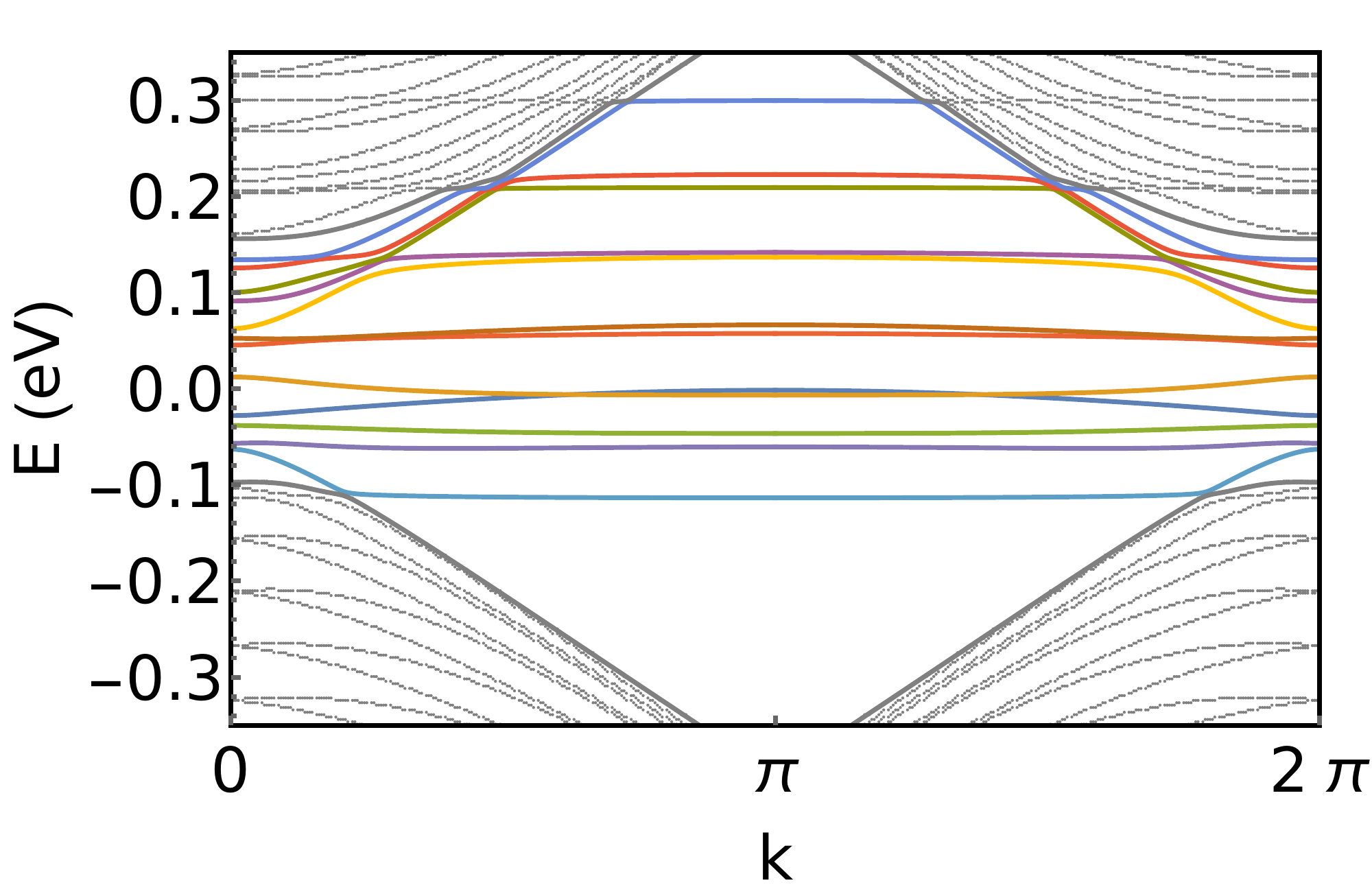}
	\includegraphics[width=0.3\textwidth]{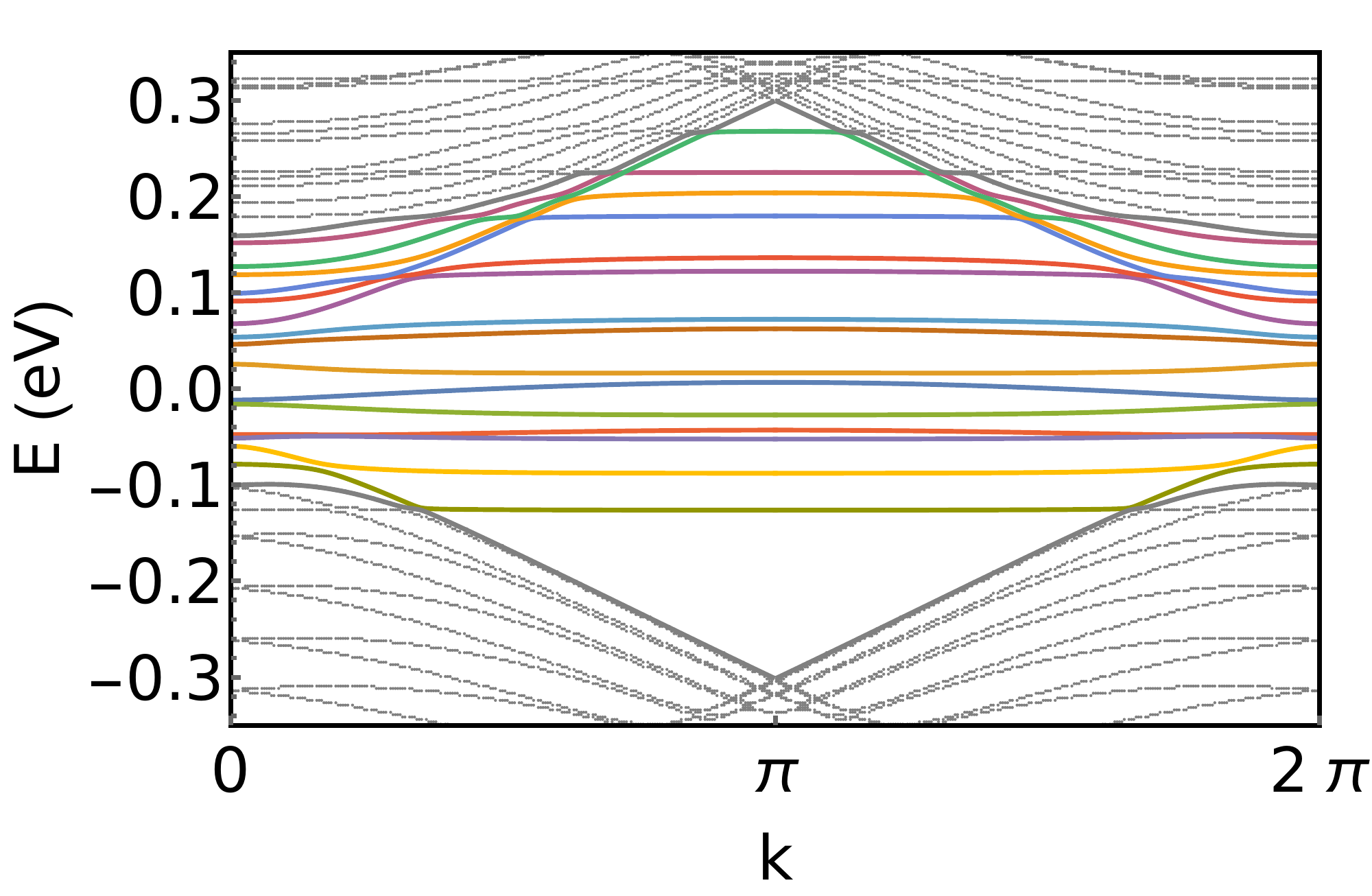}
	\caption{Band structures for a nanoribbon 144 lattice sites in width, parameters given in Eq. \ref{eq:parameters} except for $V_s^e$ = 100 meV. From left to right, supercells are 6 $\times$ 6, 18 $\times$ 18 and 24 $\times$ 24 graphene unit cells respectively; results for 12 $\times$ 12 and 48 $\times$ 48 cases are given in Fig. \ref{fig:3x48}.}\label{fig:scsize}
\end{figure}

We can directly investigate the effect of changing supercell size on the flat bands using the tight-binding model described in Section \ref{sec:nanoribbon}. Considering nanoribbons 144 graphene unit cells in width as before, we calculate the band structure for increasingly large supercells using the same superlattice potentials as the strong even scalar potential case in Fig.~\ref{fig:3x48}. Results are given in Fig. \ref{fig:scsize}, larger supercells producing an increasing number of bands within the bulk gap, eventually filling the gap for realistically large supercells as discussed in the main text and Fig.~\ref{fig:3x48}. In addition, the gap between bulk bands (most notably at $k=\pi$) tends towards the value calculated using the continuum model. These results are qualitatively similar regardless of our choice of superlattice parameters.

We attribute this behaviour to the folding of our graphene FBZ due to the superlattice: bands across the graphene FBZ are mapped on to the smaller superlattice FBZ, so accordingly the number of bands in the superlattice FBZ will be larger dependent on the size of the superlattice. Without superlattice perturbation parameters the flat band hosted on a zigzag edge of graphene will be mapped on to itself, producing degenerate subbands in the superlattice FBZ. Finite superlattice perturbations alter the energies of these bands, lifting the degeneracy and filling an average bulk gap. Note that while a $3n \times 3n$ supercell results in $4n$ flat bands, as stated in the main text, our tight-binding model only considers a single valley so results shown will host $2n$ flat bands.

In the continuum model outlined in Section \ref{sec:continuum}, changing the supercell size changes the magnitude of the superlattice reciprocal lattice vectors $|\vec{G}_n|$; this is inversely proportional to the number of graphene unit cells included in a single supercell, larger supercells corresponding to a smaller superlattice FBZ. Using the method described in Sec.~\ref{sec:Chern} we recalculate the Chern numbers for the first four valence and conductions subbands, labeled relative to zero energy ($-4 \dots 4$ respectively), for supercells of varying size $N_{SL} \times N_{SL}$ and the superlattice parameters in Eq. \ref{eq:parameters}. As outlined in the main text, provided the Chern numbers of occupied subbands are non-trivial the system remains a valley Hall insulator. Results found using 10 plaquettes to calculate the Chern number for varying $N_{SL}$ are given in Table~\ref{tab:Chern}.

\begin{table}
	\caption{Chern numbers for the conduction and valence band as a function of supercell size.}\label{tab:Chern}
 \begin{center}
\begin{tabular}{ |c| c c c c c| }
 \hline \diaghead{Bands $NSL$}{Band}{$N_{SL}$} & 20 & 30 & 40 & 50 & 60 \\ \hline 
 $-4$ & 1 & -2 & 1 & 1 & -2 \\  
 $-3$ & -1 & 2 & -1 & -1 & 3 \\
 $-2$ & 0 & 0 & 1 & 1 & 0 \\
 $-1$ & 1 & 1 & 0 & 0 & 0 \\
 $1$ & -1 & -1 & -1 & 0 & 0 \\
 $2$ & 0 & 0 & 2 & 1 & 1 \\
 $3$ & -3 & -2 & 0 & 0 & 0 \\
 $4$ & 4 & 3 & -1 & -1 & -1 \\
 \hline	
	
\end{tabular} \end{center}
\end{table}

\section{Current Operator}
We shall look at the equilibrium expectation value of the current operator based on our knowledge of the wavefunctions of charge carriers in a nanoribbon. 
The wavefunctions are all of the discrete form
\begin{equation}
\psi_{l,k_x}(\vec r_{i,j}),
\end{equation}
with energy $\epsilon_l(k_x)$ for a given band $l$. Here $i,j$ labels the lattice sites, and $\vec r$ is the coordinate-space position of the atom labeled by $(i,j)$ inside the nanoribbon. The quantity $k_x$ is the electron momentum parallel to the ribbon.

The current operator is proportional to the velocity operator, which in the well-known case of a 1D lattice model may be defined as \cite{KAL11}
\begin{equation}
j_{1D} = -i w \left(c^\dagger_{i}c_{i+1}-c^\dagger_{i+1}c_{i}\right).
\end{equation}
We similarly define a current operator for a 2D nanoribbon. We assume an applied voltage parallel to the nanoribbon, perturbing the system by introducing a preferential hopping direction, and ignore
the small effects of the periodic modulations so that there is no transverse current flowing perpendicular to the edges. The electrons can thus only hop between nearest neighbors along a series of 1D zigzag chains. This allows us to define the current at the vertical position of each, which we choose as the average vertical displacement of the bonds in each chain. For a bond oriented in direction $\delta\vec{r}=\vec r_{i'j'}-\vec r_{ij}$ the current operator is therefore
\begin{equation}
\left(j_x\right)_{\vec{r},\vec{r}'} = j_x(\vec r, \delta\vec r)\delta_{|\delta\vec r|,1} 
\propto i\,  \delta r_x\,\delta_{|\delta\vec r|,1}.
\end{equation}

In order to measure the current's vertical dependence across the width of our nanoribbon we introduce an operator that gives the average vertical displacement of each bond, $P_y = \delta_{(r_y + r_y')/2-y}\delta_{|\delta\vec r|,1}$. As we have mirror symmetry about any axis perpendicular to the ribbon direction, however, the expectation value of $P_y$ will be zero -- without incorporating the effect of the external voltage, electrons are equally likely to hop in both directions. At equilibrium the current contributions from $\pm k_x$ will be equal and opposite, and we must include a perturbation of the states to produce a net current.

We achieve this by decomposing our net current into left- ($-x$) and right-moving ($x$) components. Results shown both here and in the main text have been calculated for the range $\pi \leq k_x \leq 2\pi$ such that dispersive bands have positive dispersion; we have verified that the corresponding current from $0 \leq k_x \leq \pi$ is both equal in magnitude and flows in the opposite direction. This corresponds to the assumption that applying a small voltage generates a shift in the chemical potential, providing an imbalance between left- and right-moving currents. 

We therefore calculate the expectation value of $j_x$ at fixed energy $E$ at the Fermi energy:
\begin{align}\label{eq:current}
J_+(E,y )&=\int_\pi^{2\pi}  d k_x\sum_l \delta(\epsilon_l(k_x)-E)\sum_{\vec r, \vec r'}  \langle \psi_{l,k_x} | \vec r\rangle
(j_x)_{\vec r, \vec r'} \delta_{( r_y+ r'_y)/2-y} \langle \vec r'|\psi_{l,k_x}\rangle
\nonumber\\
&\propto \int_\pi^{2\pi}  d k_x\sum_l \delta(\epsilon_l(k_x)-E) \sum_{\vec r, \vec r'} \langle \psi_{l,k_x} | \vec r\rangle
i  \delta r_x\,\delta_{|\delta\vec r|,1} \delta_{( r_y+ r'_y)/2-y} \langle \vec r'|\psi_{l,k_x}\rangle.
\end{align}

We further exploit the reflection symmetry of our nanoribbon to enforce an overall positive current contribution from each band. It is possible that the current contribution from a given wavefunction will have counter-propagating components along the width of our nanoribbon: to ensure that the current produced by each band travels in the same direction overall, we multiply each current as calculated in Eq. \ref{eq:current} by its overall sign calculated across the nanoribbon's width.

\begin{figure}
	\includegraphics[width=0.32\textwidth]{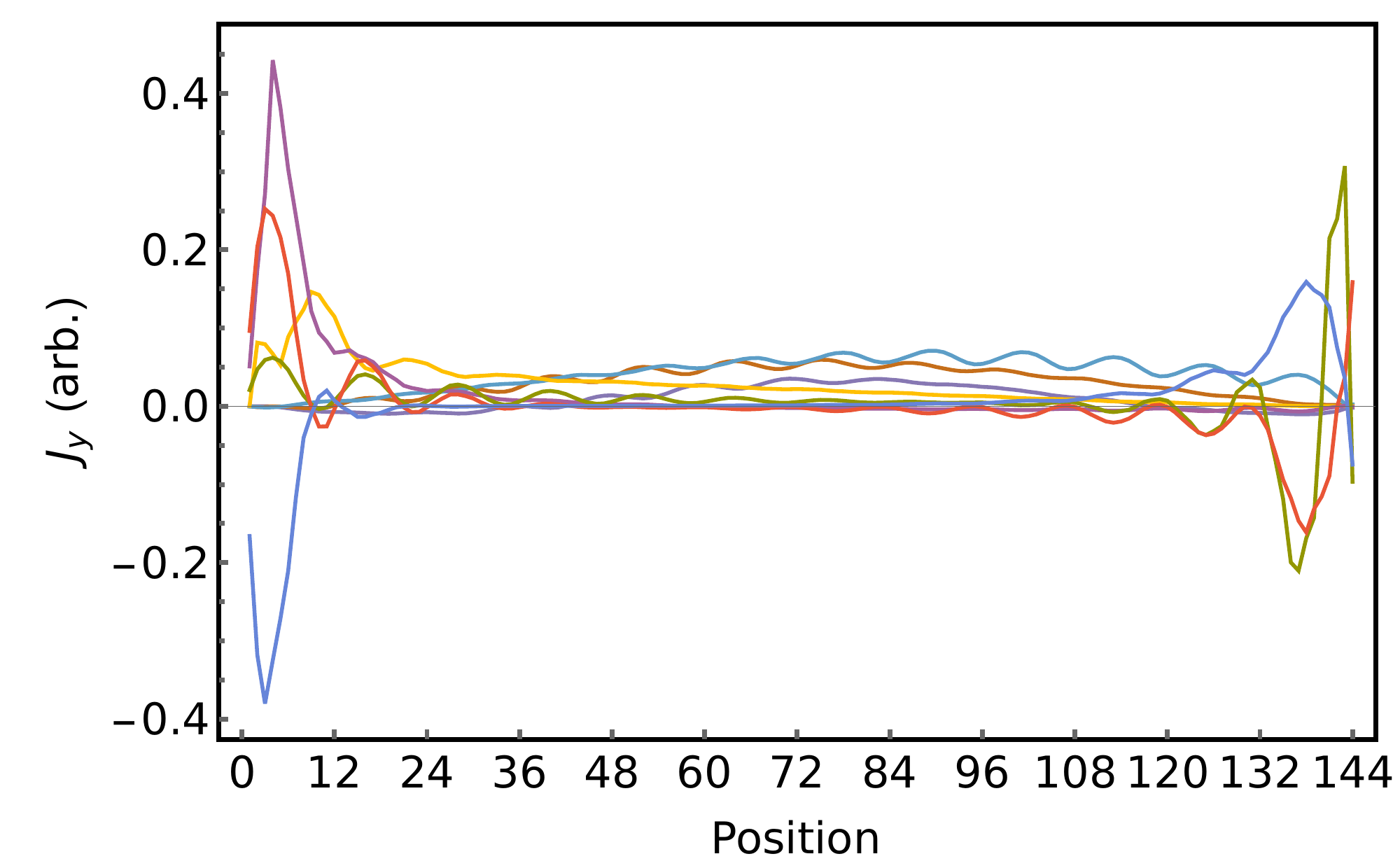}
	\includegraphics[width=0.32\textwidth]{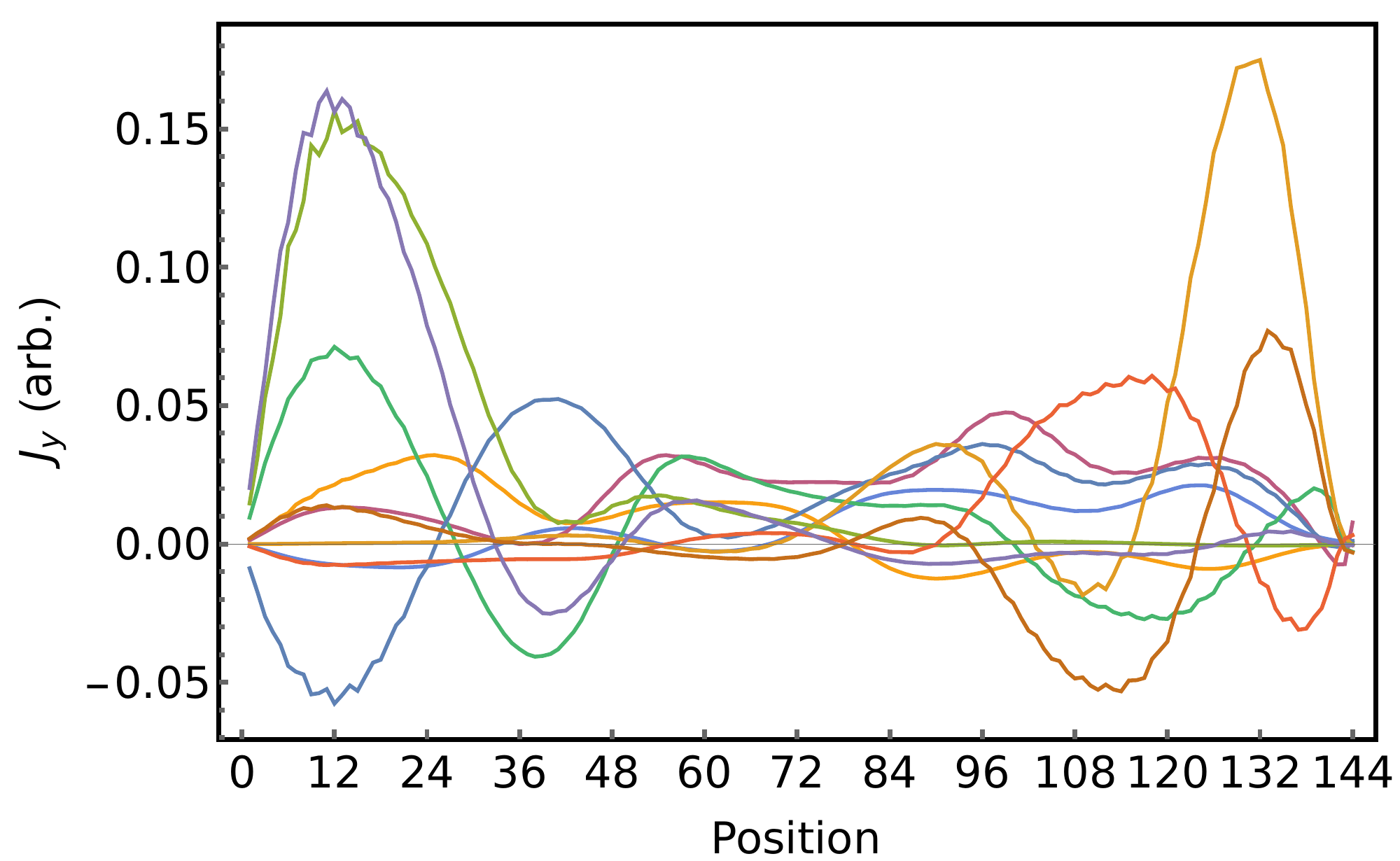}
	\includegraphics[width=0.32\textwidth]{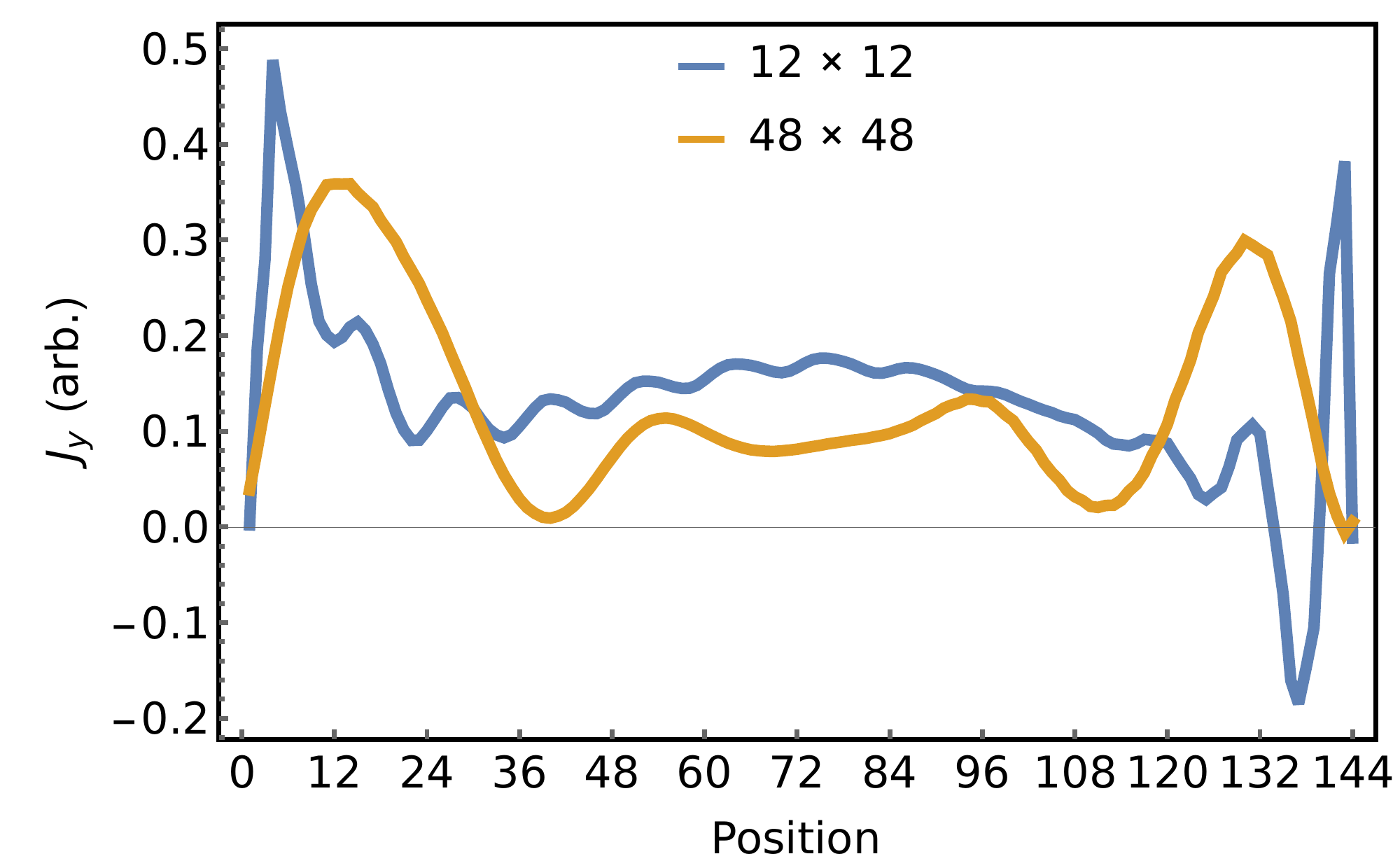}
	\caption{Currents across a nanoribbon 144 lattice sites in width due to a superlattice with our given parameters and a larger even scalar potential, $V_s^e$ = 0.1 eV. Contributions from individual wavefunctions are given for a superlattice with a unit cell of 12 $\times$ 12 graphene unit cells, -0.1 $\leq$ E $\leq$ 0.1 eV (\textit{left}) and 48 $\times$ 48 graphene unit cells, -0.05 $\leq$ E $\leq$ 0.05 eV (\textit{center}). The total current in each case is also given (\textit{right}).}\label{fig:sjscurrent}
\end{figure}

\begin{figure}
	\includegraphics[width=0.32\textwidth,valign=t]{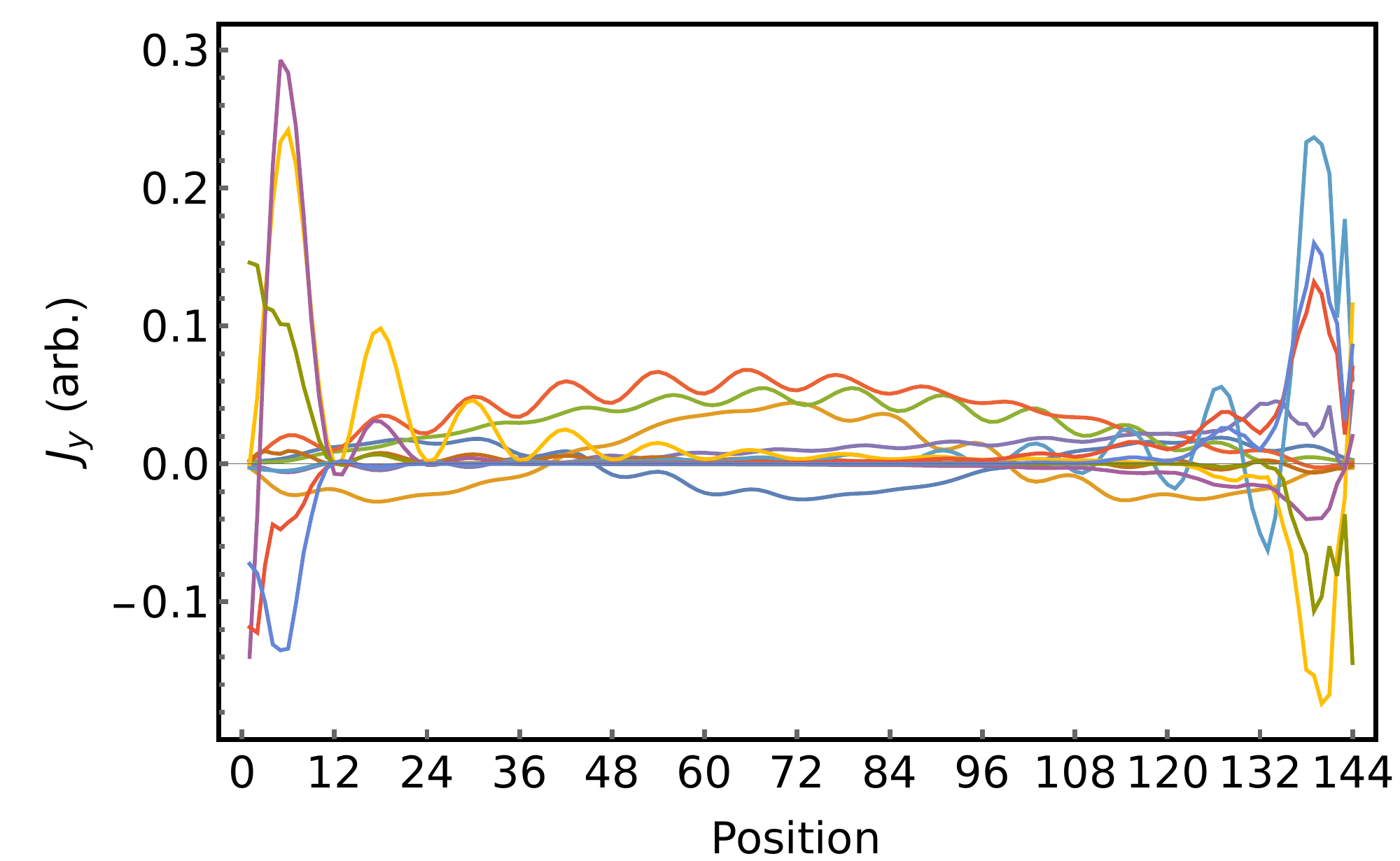}
	\includegraphics[width=0.32\textwidth,valign=t]{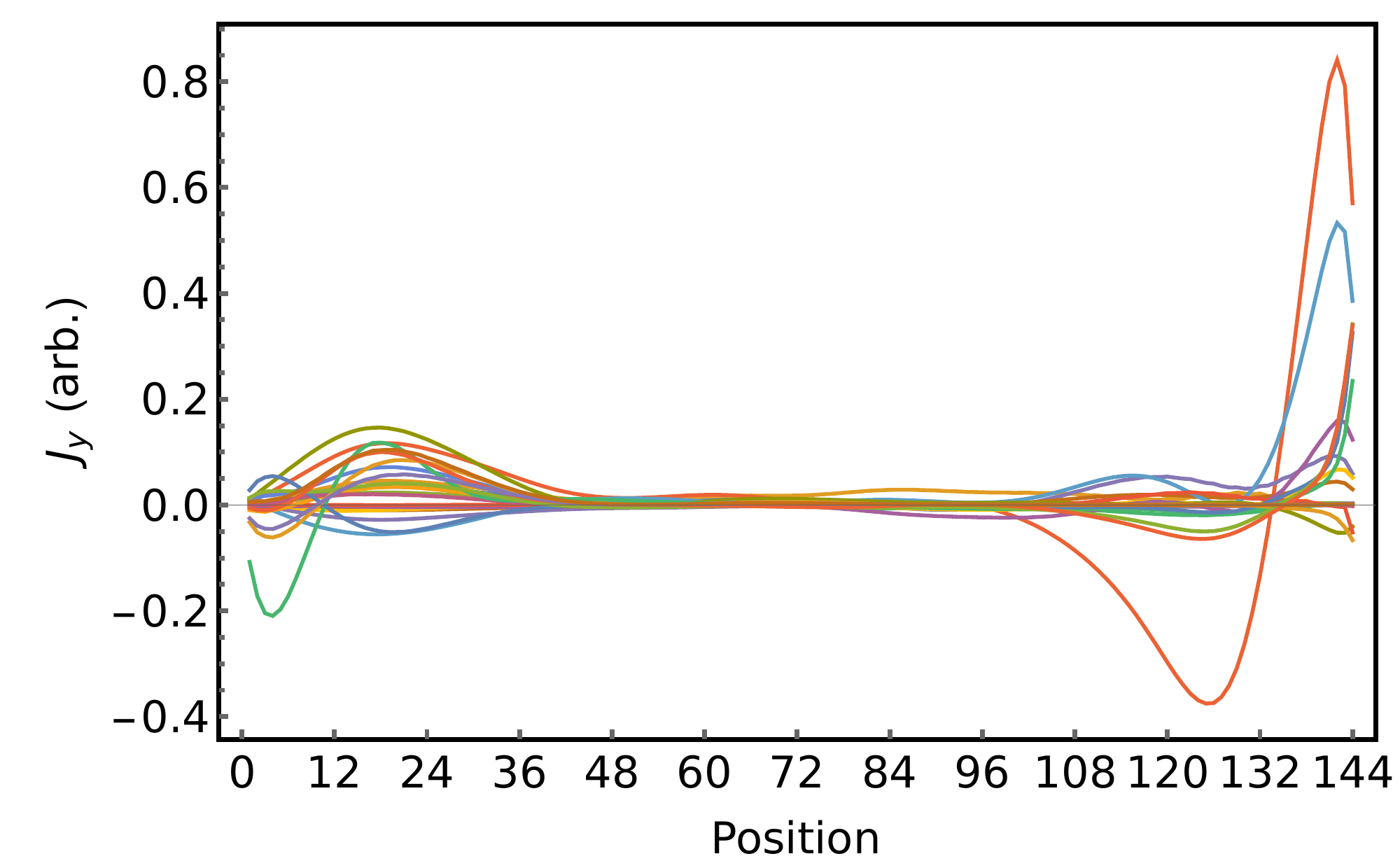}
	\includegraphics[width=0.34\textwidth,valign=t]{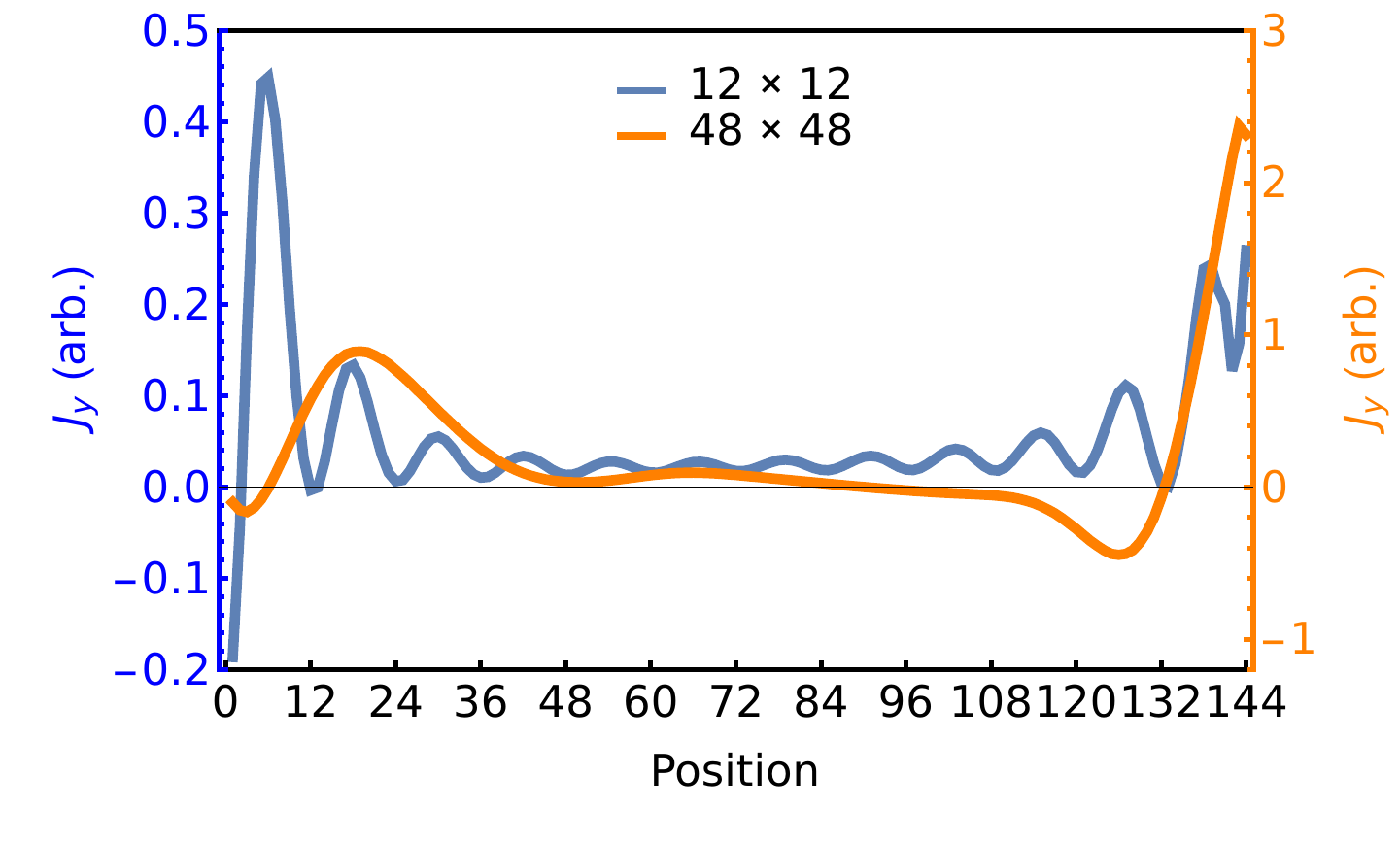}
	\caption{Currents across a nanoribbon 144 lattice sites in width due to a superlattice with our given parameters and a larger odd scalar potential, $V_s^e$ = 0.1 eV. Contributions from individual wavefunctions are given for a superlattice with a unit cell of 12 $\times$ 12 graphene unit cells, -0.1 $\leq$ E $\leq$ 0.1 eV (\textit{left}) and 48 $\times$ 48 graphene unit cells, -0.05 $\leq$ E $\leq$ 0.05 eV (\textit{center}). The total current in each case is also given (\textit{right}).}\label{fig:sjacurrent}
\end{figure}

Since the application of a voltage across the ribbon means that the Fermi energy varies along its length, we can not work with a single state. As we only have equilibrium results, the best we can do is to take an average of the current operator over a range of energies. Thus, we find the total current due to our flat bands by summing over the current contributions from all bands $l$ within a given energy range; we add all currents calculated provided $E_1 \leq \epsilon_l \leq E_2$, lower and upper energies $E_1$ and $E_2$ chosen such that as far as is possible only flat bands are included. The total current and each wavefunction's contribution for the superlattice parameters used in Fig.~\ref{fig:3x48} is given in Figs.~\ref{fig:sjscurrent} and \ref{fig:sjacurrent}: energy ranges for different superlattice cell sizes have been set such that current contributions from flat bands are not hidden by those from bulk bands. We find that our model does produce  edge transport, though it may be suppressed by bulk behavior depending on the energy range used.

Here we highlight a possible ambiguity concerning the sign of our model current operator: as can be seen in both Figs. \ref{fig:sjscurrent} and \ref{fig:sjacurrent}, enforcing a net positive current per wavefunction over the entire width of the nanoribbon may lead to counterpropagating edge modes if the bulk contribution
of a mode is almost equal but opposite to the edge one. It follows that we may therefore incorrectly estimate the proportion of edge transport in our nanoribbon; to resolve this issue one should perform more detailed calculations using non-equilibrium methods to calculate the current.

\end{document}